\documentclass[a4paper,12pt]{iopart}
\expandafter\let\csname equation*\endcsname\relax
\expandafter\let\csname endequation*\endcsname\relax
\usepackage{amsmath}

\usepackage{afterpage}
\usepackage{pdfpages}
\usepackage{booktabs}
\usepackage{xcolor}

\usepackage[numbers,square,sort&compress]{natbib} 
\usepackage[colorlinks=true, linkcolor=blue, citecolor=blue, urlcolor=blue]{hyperref}
\usepackage{subfig} 
\usepackage{graphicx} 
\usepackage[format=plain,justification=justified]{caption} 
\usepackage{orcidlink}

\begin{document}

% ---------------------
% TITLE
% ---------------------
\title[Stacking-Directed Polarization in MoS$_2$/MoSe$_2$ Heterostructures]%
{Stacking-Directed Polarization and Excitonic Engineering in MoS$_2$/MoSe$_2$ van der Waals Heterostructures}

% ---------------------
% AUTHORS
% ---------------------
\author{
Mohammed El Amine Miloudi\,\orcidlink{0000-0002-2654-8758},
Oliver K\"uhn\,\orcidlink{0000-0002-5132-2961}
}

% ---------------------
% AFFILIATION
% ---------------------
\address{
Institute of Physics, University of Rostock, 
Albert-Einstein-Str. 23--24, 
D-18059 Rostock, Germany
}

% ---------------------
% EMAIL
% ---------------------
\ead{oliver.kuehn@uni-rostock.de}

% ---------------------
% DATE (OPTIONAL)
% ---------------------
\vspace{10pt}
\begin{indented}
\item[]\today
\end{indented}
% ---------------------
% ABSTRACT
% ---------------------
\begin{abstract}
The stacking-dependent polarization and excitonic response of MoS$_2$/MoSe$_2$ heterostructures were investigated using GW+BSE many-body perturbation theory. While homobilayer MoS$_2$ exhibited a switchable interlayer dipole driven by registry-induced symmetry breaking, the MoS$_2$/MoSe$_2$ hetero-interface remained pinned by the intrinsic chemical potential mismatch between sulfur and selenium. In 2L-MoS$_2$/MoSe$_2$ trilayers, the stacking sequence enabled a deterministic control of photogenerated electrons between the central and bottom MoS$_2$ layers, governed by internal electric fields and quasiparticle band-edge shifts of 60--70~meV. Our calculations predicted a 36~meV interlayer excitonic shift, in remarkable agreement with recent experiments. These results elucidate the microscopic link between atomic registry and many-body interactions, establishing transition metal dichalcogenide trilayers as a potential platform for sliding ferroelectricity and programmable optoelectronic functionalities.
\end{abstract}

% ---------------------
% INTRODUCTION
% --------------------
\section{Introduction}

Two-dimensional (2D) van der Waals (vdW) heterostructures, formed by stacking atomically thin crystals via weak interlayer interactions, have emerged as highly versatile platforms for exploring quantum phenomena and engineering optoelectronic functionalities beyond what bulk materials can offer \cite{zhao22_3418, xue21_021316, viznerstern21_1462, chen25_10, niu23_5578, li24_42501, tran21_022002}. The vdW nature of these interactions facilitates precise control over stacking sequence, rotational alignment, and interlayer spacing, giving rise to Moiré superlattices and enabling atomic-scale tuning of electronic band structures, excitonic resonances, and interfacial polarization. These properties make vdW heterostructures ideal for manipulating light–matter interactions at the atomic scale.

Among 2D materials, transition metal dichalcogenides (TMDs) such as MoS$_2$, MoSe$_2$, WS$_2$, and WSe$_2$ have attracted particular attention due to their direct band gaps in the visible--near-infrared range, strong spin--orbit coupling, and tightly bound excitons with binding energies of 0.5--0.9~eV~\cite{amin15_075439, berkelbach13_045318, mak10_136805, wang17_44712, xiao12_196802}. These high binding energies, nearly an order of magnitude larger than in conventional bulk semiconductors, arise from reduced dielectric screening and enhanced Coulomb interactions in two dimensions. Consequently, monolayer TMDs exhibit exceptionally strong light absorption, up to 5--10\% of incident light per layer, and robust valley-selective optical transitions, making them ideal candidates for excitonic and valleytronic applications~\cite{mak10_136805, wang17_44712, wang18_021001}.

When stacked into heterostructures, TMD monolayers often display type-II band alignment, which spatially separates electrons and holes and stabilizes interlayer excitons with large out-of-plane dipole moments and extended lifetimes~\cite{tran21_022002, wang17_44712}. For instance, in MoS$_2$-based heterostructures, conduction- and valence-band offsets of 0.3–0.6~eV drive ultrafast interlayer charge transfer on sub-100~fs timescales~\cite{hong14_682, ceballos14_12717}. The spatial separation of carriers suppresses radiative recombination and renders interlayer excitons highly sensitive to interfacial asymmetry and external fields, providing versatile opportunities for optoelectronic and quantum-functional  devices.

The electronic and excitonic properties of vdW heterostructures are strongly dependent on interlayer arrangement, atomic registry, and stacking symmetry. Variations in stacking alter interlayer orbital hybridization, local electrostatics, and charge distribution, which in turn modify band edges, exciton binding energies, and optical selection rules. In twisted heterobilayers, small relative rotations produce long-period Moiré superlattices that confine electrons, holes, and interlayer excitons, flatten excitonic minibands, and enable discrete exciton arrays with correlated phases and single-photon emission~\cite{tran19_71, zhang18_7651, choi21_047401, jiang21_72b, li24_42501a}. For twist angles below approximately $2^\circ$, Moiré periods can exceed 10–15~nm, generating deep periodic potentials that strongly influence exciton localization and recombination dynamics.

Interlayer excitons in TMD heterobilayers exhibit lifetimes that are orders of magnitude longer than those of intralayer excitons, reflecting their spatially indirect character. For example, vertically stacked MoSe$_2$–WSe$_2$ heterostructures show interlayer exciton lifetimes of approximately 1.8~ns at low temperature, compared to tens of picoseconds for intralayer excitons~\cite{jung22_4543}. Moiré engineering at small twist angles can further extend lifetimes into the tens-of-nanoseconds range by suppressing nonradiative decay pathways~\cite{lin24_8762b}. These long-lived excitons, combined with strong binding energies, underpin robust light--matter interactions and highlight the potential of vdW heterostructures for excitonic devices.

While Moiré superlattices provide a static landscape for particle localization, lateral interlayer sliding offers a continuous and dynamical pathway to sample the stacking configuration space without changing the Moiré period \cite{zhang25_76, cao22_57492}. By systematically shifting the stacking configuration, sliding can reversibly modulate interfacial polarization, band topology, charge transfer, and exciton energies. This registry-dependent control complements Moiré engineering, enabling the design of heterostructures with electrically controllable dipoles and tailored excitonic landscapes \cite{pan25_126603, yang24_1389b}.

The interplay between interlayer polarization and excitonic dynamics opens additional avenues for quantum-functional design. In inversion-asymmetric or 3R-stacked TMD heterostructures, spontaneous out-of-plane polarizations on the order of 0.1–0.3~C~m$^{-2}$ produce internal fields up to 0.1–0.5~V~nm$^{-1}$, sufficient to shift band edges, redistribute charge, and tune interlayer exciton energies~\cite{jiang21_72, hou25_34954}. Controlled stacking and polarization engineering thus enable electrically programmable excitonic landscapes, ferroelectric switching, and multifunctional optoelectronic behavior \cite{schwandt-krause26_214}.

Despite these advances, a systematic understanding of how atomic registry governs many-body excitonic behavior in MoS$_2$-based heterostructures remains elusive. Existing theoretical frameworks often rely on mean-field approximations that fail to account for the non-local screening and enhanced Coulomb interactions inherent to these 2D interfaces. Consequently, the fundamental relationship between local stacking order and the resulting interfacial dipoles particularly at the many-body GW--BSE level requires in-depth elucidation. This motivates a rigorous computational study of the electronic and excitonic landscapes in these systems.

Here, we employ  a high-fidelity theoretical framework to investigate stacking-dependent polarization and excitonic properties in MoS$_2$-based vdW heterostructures, building upon recent experimental demonstrations of ferroelectric control in 3R-stacked systems \cite{schwandt-krause26_214}. Utilizing density functional theory (DFT) with hybrid functionals and many-body perturbation theory within the EVGW$_0$ (eigenvalue-self-consistent GW$_0$)–BSE framework, we move beyond standard approximations to precisely quantify how atomic registry and interlayer asymmetry govern interfacial dipoles and quasiparticle band alignments. We demonstrate that AB and BA registries generate interfacial dipoles of opposite polarity, resulting in switchable band-edge shifts while maintaining the robust type-II alignment necessary for charge separation. Furthermore, we show that in trilayer configurations, stacking-induced asymmetry allows for the fine-tuning of polarization states and excitonic transitions, offering a predictive road map for the design of non-volatile excitonic transistors and atomically controlled ferroelectric optoelectronic logic.

% ---------------------
% METHODS
% ---------------------
\section{Computational Framework}

The structural and electronic properties of MoS$_2$/MoSe$_2$ heterostructures were investigated within the framework of DFT, as implemented in the Vienna \emph{Ab initio} Simulation Package (VASP)~\cite{blochl94_17953a, kresse99_1758, kresse96_11169a, kresse96_15a}. Electron--ion interactions were described using the projector-augmented wave (PAW) formalism, with valence configurations of Mo ($4s^2 4p^6 4d^5 5s^1$), S ($3s^2 3p^4$), and Se ($4s^2 4p^4$). Exchange--correlation effects were treated within the generalized gradient approximation using the Perdew--Burke--Ernzerhof functional~\cite{perdew96_3865a}, supplemented by Grimme’s D3 scheme to account for long-range dispersion interactions.

All calculations were performed using a plane-wave kinetic-energy cutoff of 450~eV and $\Gamma$-centered Monkhorst--Pack $k$-point meshes of $12 \times 12 \times 1$. A vacuum region exceeding 20~\AA\ was introduced along the out-of-plane direction to suppress spurious interactions between periodic replicas. Atomic coordinates were fully optimized until residual forces were below 0.01~eV~\AA$^{-1}$ and total-energy variations were smaller than $10^{-8}$~eV. Systematic convergence tests with respect to the $k$-point grid (ranging from $7 \times 7 \times 1$ to $20 \times 20 \times 1$) confirm that total energy differences remain within less than 0.3~meV per unit cell from $9 \times 9 \times 1$ onward, validating our production grid selection (see Supplemental Material (Suppl. Mat.), Fig.~S1). 

Commensurate heterostructure models were constructed by first optimizing isolated MoS$_2$ and MoSe$_2$ monolayers. 
%The optimized lattice constants, along with the corresponding out-of-plane parameters for bilayer and trilayer stackings, are summarized in Table S1. 
Monolayers of MoS$_2$ and MoSe$_2$ were found to have lattice constants of 3.166~\AA\ and 3.295~\AA, respectively. For bilayers and heterobilayers, AA, AB, and BA stacking configurations were considered, while multiple trilayer stackings (AAA, ABA, ABB, BAA, BAB, AAB, and BBA) were systematically analyzed. Lattice matching in heterobilayers was achieved by adopting a common in-plane lattice parameter, which corresponds to biaxial strains of +2.1\% in MoS$_2$ and -1.9\% in MoSe$_2$. While such strain magnitudes can quantitatively shift absolute band edges and excitonic transition energies, the relative energetic trends and stability hierarchies among competing local stacking registries remain robust, as all compared configurations experience identical uniform strain fields within the rigid $1 \times 1$ commensurate model. Furthermore, while experimental samples with small lattice mismatches typically form long-period moiré superlattices accompanied by local atomic and strain relaxation, our commensurate $1 \times 1$ framework provides fundamental insights into the local stacking-dependent potential energy landscape and intrinsic energetic hierarchies governing these van der Waals heterostructures. 

Electronic band structures and band gaps were evaluated at multiple theoretical levels. DFT calculations were performed using PBE, including spin--orbit coupling (SOC) where indicated, as well as the screened hybrid HSE06 functional without SOC. Many-body effects beyond DFT were incorporated through EVGW$_0$@PBE~\cite{hybertsen86_5390b, wilhelm21_1662}, in which quasiparticle energies in the Green’s function were iteratively updated while the orbitals and screened Coulomb interaction were kept fixed at the PBE level. Excitonic effects in the optical response were explicitly treated by solving the Bethe--Salpeter equation (BSE)~\cite{hanke1980_prb, strinati1988_rnc}.

All EVGW$_0$ and BSE calculations employed PAW datasets optimized for many-body perturbation theory, including semicore states, together with carefully converged $k$-point meshes and numbers of unoccupied bands. Bilayer MoS$_2$ and MoS$_2$/MoSe$_2$ heterobilayer were computed using a $\Gamma$-centered $12 \times 12 \times 1$ $k$-point grid with 32 unoccupied bands, while the 2L-MoS$_2$/MoSe$_2$ systems used a $\Gamma$-centered $9 \times 9 \times 1$ grid with 64 unoccupied bands. All numerical parameters were systematically tested to ensure the reliability of quasiparticle energies and excitonic spectra.
%
% ---------------------
% RESULTS AND DISCUSSION
% ---------------------
\section{Results and Discussion}
% --------------------------------------------------------------------------------------------------
\subsection{Bilayer MoS$_2$ and MoS$_2$/MoSe$_2$ Heterobilayers}
\label{sec:Bilayer MoS$_2$ and MoS$_2$/MoSe$_2$ Heterobilayers}
% --------------------------------------------------------------------------------------------------
\subsubsection{Stacking-Dependent Polarization and Interlayer Charge Redistribution}
\label{sec:Stacking-Dependent Polarization and Interlayer Charge Redistribution}\

%--------------------------------------------Figure 1 ------------------------------------------------------
\begin{figure}[ht]
\centering
\includegraphics[width=1\textwidth]{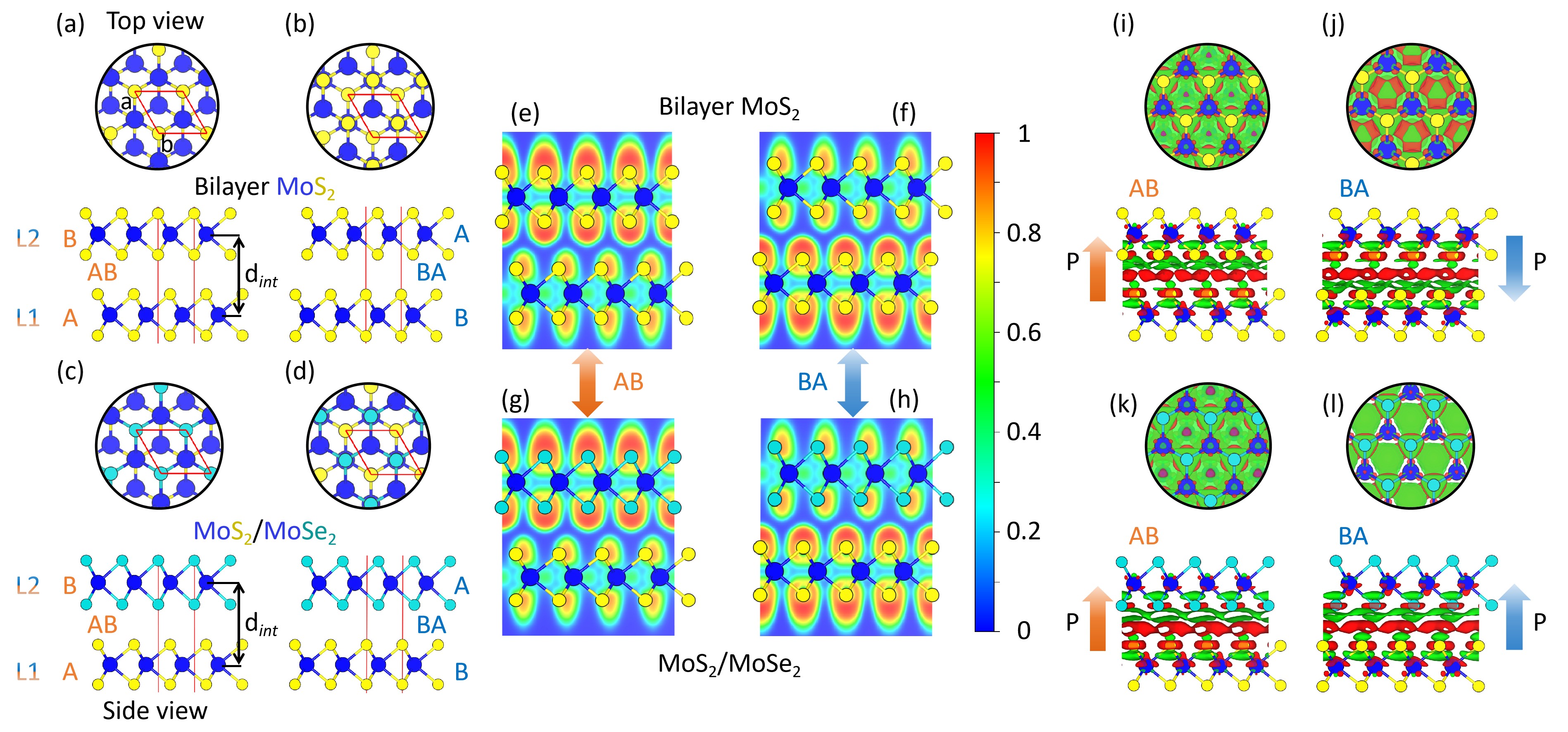}
\caption{Top and side views of the atomic configurations for (a) AB-stacked and (b) BA-stacked MoS$_2$ homobilayers, and (c) AB-stacked and (d) BA-stacked MoS$_2$/MoSe$_2$ heterobilayers. Panels (e)–(h) display the corresponding electron localization function (ELF) contour plots with an interval of 0.1, illustrating the covalent nature of intralayer bonding and the vdW gap. Panels (i)–(l) show the charge density difference ($\Delta\rho$) isosurfaces for each configuration. Red and green regions represent charge depletion and accumulation, respectively, visualized at an isovalue of 0.001 $e$/\AA$^3$. The variations in stacking order and chemical composition significantly modulate the interlayer coupling and the resulting interfacial charge redistribution.}
\label{fig:Fig1}
\end{figure}
%------------------------------------------------------------------------------------------------------------

% ---------------------------------atomic configurations-Fig. 1 (a-d)---------------------------------------------------
The impact of stacking registry on interlayer coupling in both homobilayer MoS$_2$ and MoS$_2$/MoSe$_2$ heterobilayers was systematically evaluated for AA, AB, and BA configurations. The calculated in-plane lattice constants ($a$) and out-of-plane lattice parameters ($c$) for monolayer and bilayer stackings are summarized in Table~S1, providing a quantitative basis for subsequent analysis of interlayer interactions.

Following full structural relaxation, the AA registry preserves high structural symmetry, whereas the staggered AB and BA configurations break global inversion symmetry, thereby inducing a finite out-of-plane electric polarization, $P_z$. The optimized geometries for these polar registries, including both top and side views, are illustrated for the homobilayer in Fig.~\ref{fig:Fig1}a,b and for the heterobilayer in Fig.~\ref{fig:Fig1}c,d, respectively. Corresponding views of the non-polar AA stacking are provided in Fig.~S2a,b of the Suppl. Mat.

The evolution of the lattice parameters from monolayer to bilayer provides insight into the interplay between intralayer strain, interlayer steric repulsion, and vdW interactions. As listed in Table~S1, monolayer MoS$_2$ and MoSe$_2$ exhibit equilibrium in-plane lattice constants of 3.166~\AA\ and 3.295~\AA, respectively. Upon forming a heterobilayer, the in-plane lattice constant $a$ converges to approximately 3.228~\AA, effectively representing an arithmetic mean of the constituent monolayers. This indicates a commensurate strain regime in which the energy cost of stretching MoS$_2$ is balanced by compressing MoSe$_2$, enabling commensurable stacking with minimal total elastic energy. For the homobilayer MoS$_2$, AB and BA stackings show only minor deviations from the monolayer $a$ due to symmetric strain distribution between identical layers.

The out-of-plane lattice constant $c$ is strongly influenced by stacking registry. In the AA configuration, Mo atoms and chalcogen atoms from opposing layers are vertically aligned, maximizing electron cloud overlap and inducing strong Pauli repulsion. This steric effect pushes the layers apart, resulting in the largest $c$ values and Mo--Mo interlayer separations ($d_{\mathrm{int}}$), reaching 6.74~\AA\ for homobilayer MoS$_2$ and 6.94~\AA\ for MoS$_2$/MoSe$_2$. In contrast, the staggered AB and BA configurations allow lateral atomic offsets, reducing localized electronic repulsion and enabling attractive vdW interactions to bring the layers closer to each other. In these registries, $d_{\mathrm{int}}$ decreases to 6.43~\AA\ for MoS$_2$ and 6.62~\AA\ for MoS$_2$/MoSe$_2$, with corresponding contractions in $c$. The systematic expansion of approximately 0.20~\AA\ in the heterobilayer relative to the homobilayer reflects the larger vdW radius of Se compared to S.

% ----------------------------- ELF -----------------------------
To characterize the bonding nature and spatial distribution of electrons, the electron localization function (ELF) was analyzed, as illustrated in Fig.~\ref{fig:Fig1}e--h. ELF values approaching unity indicate regions of strong covalent bonding or lone-pair character, whereas values near zero denote regions of low electron density. Across all configurations, a pronounced low-ELF region (ELF~$\approx 0$) is observed within the interlayer space, identifying the vdW gap and confirming the absence of significant orbital overlap.

The intralayer bonding characteristics are further elucidated by these ELF distributions. In the AB-stacked systems (Fig.~\ref{fig:Fig1}e,g), the first layer exhibits moderate localization around chalcogen sites, while the second layer shows stronger, more spatially extended localization, indicated by orange-to-red regions. In the BA stacking (Fig.~\ref{fig:Fig1}f,h), this behavior is inverted, demonstrating a stacking-dependent redistribution of the local electronic environment and highlighting the sensitivity of intralayer charge to the vertical atomic registry.

% ----------------------------- CDD -----------------------------
Interlayer charge redistribution was quantified using charge density difference (CDD) analysis. The three-dimensional CDD isosurfaces, illustrated in Fig.~\ref{fig:Fig1}i--l, use red and green contours to indicate charge depletion and accumulation, respectively (isovalue = $0.001~\mathrm{e}/\text{\AA}^3$). The CDD, $\Delta \rho(\mathbf{r})$, is defined as
\begin{equation}
\Delta \rho(\mathbf{r}) = \rho_{\mathrm{bilayer}}(\mathbf{r}) - \rho_{L_1}(\mathbf{r}) - \rho_{L_2}(\mathbf{r}),
\end{equation}
where $\rho_{\mathrm{bilayer}}$, $\rho_{L_1}$, and $\rho_{L_2}$ are the charge densities of the bilayer and isolated monolayers at the same atomic coordinates. The planar-averaged redistribution perpendicular to the interface (along the $z$-axis) is
\begin{equation}
\Delta \rho(z) = \frac{1}{A} \int \Delta \rho(x,y,z)\, dx\, dy,
\end{equation}
where $A$ denotes the in-plane unit-cell area, yielding units of $\mathrm{e}/\text{\AA}^3$ for $\Delta \rho(z)$. The out-of-plane macroscopic electric polarization $P_z$ (in units of $\mathrm{C/m}$) was evaluated directly for the two-dimensional unit cell using the Berry phase framework.

The stacking registry dictates the electronic landscape of the bilayers through a competition between structural symmetry and local orbital hybridization. In AA-stacked bilayers, the preservation of an interlayer mirror symmetry plane ($\sigma_z$) enforces a symmetric charge redistribution, $\Delta \rho(z) = \Delta \rho(-z)$, fundamentally precluding an out-of-plane dipole ($P_z = 0$). This is corroborated by the symmetry-protected CDD isosurfaces (Fig.~S2a,b) and planar-averaged profiles (Fig.~S2c,d), which exhibit distributions centered symmetrically across the vdW gap. The vertical alignment of Mo atoms and opposing chalcogen planes in layers L1 and L2 maintains a balanced interlayer hybridization, preserving a non-polar, purely vdW-mediated interaction.

\begin{figure}[ht]
\centering
\includegraphics[width=1\textwidth]{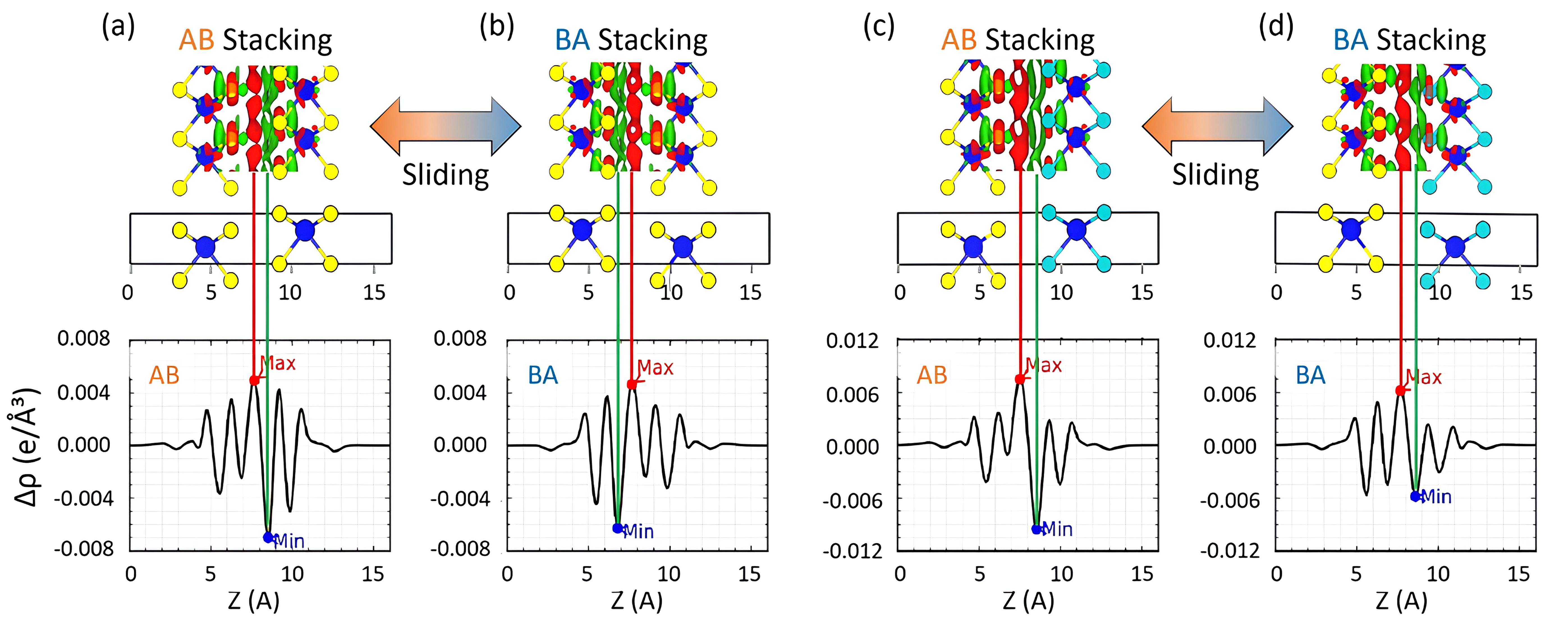}
\caption{Planar-averaged charge density difference ($\Delta\rho$) profiles for MoS$_2$ and MoS$_2$/MoSe$_2$ bilayers. Panels (a)–(d) display the out-of-plane ($z$) distribution of charge for AB- and BA-stacked configurations. The top insets map the 3D $\Delta\rho$ isosurfaces to the 1D profiles; red and green vertical lines correlate the quantitative maxima (Max) and minima (Min) with physical regions of electron accumulation and depletion, respectively. The sequence of these extrema (Min$\rightarrow$Max vs. Max$\rightarrow$Min) delineates the interfacial dipole orientation, illustrating how lateral sliding between AB and BA stackings reverses or modulates the internal electric field and interlayer coupling.}
\label{fig:Fig1-1}
\end{figure}

In contrast, the staggered AB and BA registries facilitate a pronounced, asymmetric interlayer charge transfer. For the MoS$_2$ homobilayer in the AB configuration (Fig.~\ref{fig:Fig1}i and Fig.~\ref{fig:Fig1-1}a), a spontaneous electronic redistribution occurs, with charge depletion (holes) localized near the lower layer (L1, red region) and charge accumulation (electrons) concentrated near the top layer (L2, green region). This spatial separation of charge establishes a net upward-oriented electric polarization, $P_z$ ($\uparrow$). The redistribution originates from the broken global inversion symmetry inherent to staggered registries, which permits asymmetric hybridization between the metal Mo-$4d_{z^2}$ and chalcogen S-$3p_z$ orbitals. This effect is quantified by the planar-averaged $\Delta \rho(z)$, which exhibits extrema of $+0.005$ and $-0.007~\mathrm{e}/\text{\AA}^3$. A structural transition to the BA configuration (Fig.~\ref{fig:Fig1}j and Fig.~\ref{fig:Fig1-1}b) effectively inverts this electronic profile, reversing the dipole orientation to $P_z$ ($\downarrow$). The existence of these two energetically degenerate states with opposing dipole moments confirms a switchable, bistable interlayer configuration, representing the essential microscopic prerequisite for sliding ferroelectricity in MoS$_2$ bilayers.

For the chemically asymmetric $\mathrm{MoS}_2/\mathrm{MoSe}_2$ heterobilayer (Fig.~\ref{fig:Fig1}k,l), localized electronic redistribution occurs due to the electronegativity mismatch between sulfur and selenium, exhibiting charge depletion on the $\mathrm{MoS}_2$ side and accumulation on the $\mathrm{MoSe}_2$ side. However, quantitative evaluation via the Berry phase method yields a negligible net polarization value of $P_z \approx 5\times10^{-14}~\mathrm{C/m}$, substantially smaller than the switchable homobilayer polarization of $\sim7.94\times10^{-10}~\mathrm{C/m}$. This highlights a crucial physical distinction: while local CDD isosurfaces capture microscopic orbital asymmetry and bonding variations at the interface, they do not inherently translate into a finite macroscopic dipole moment after full unit-cell integration. Consequently, the planar-averaged electrostatic potential across the heterobilayer reveals an approximately vanishing potential offset ($\Delta V_{\mathrm{hetero}} \approx 0$, $E_z \sim 0$). This demonstrates that local asymmetric charge transfer is counterbalanced by the intrinsic chemical potential profile, chemically pinning the electronic structure and suppressing switchable out-of-plane ferroelectricity.

%--------------------------------------------Figure 1-1 ------------------------------------------------------
%--------------------------------------------Figure 1-1 ------------------------------------------------------

%--------------------------------------------------------------------------------------------------------------------------
%--------------------------------------------Figure 2 ------------------------------------------------------

\begin{figure*}[ht!]
\centering
\includegraphics[width=\textwidth]{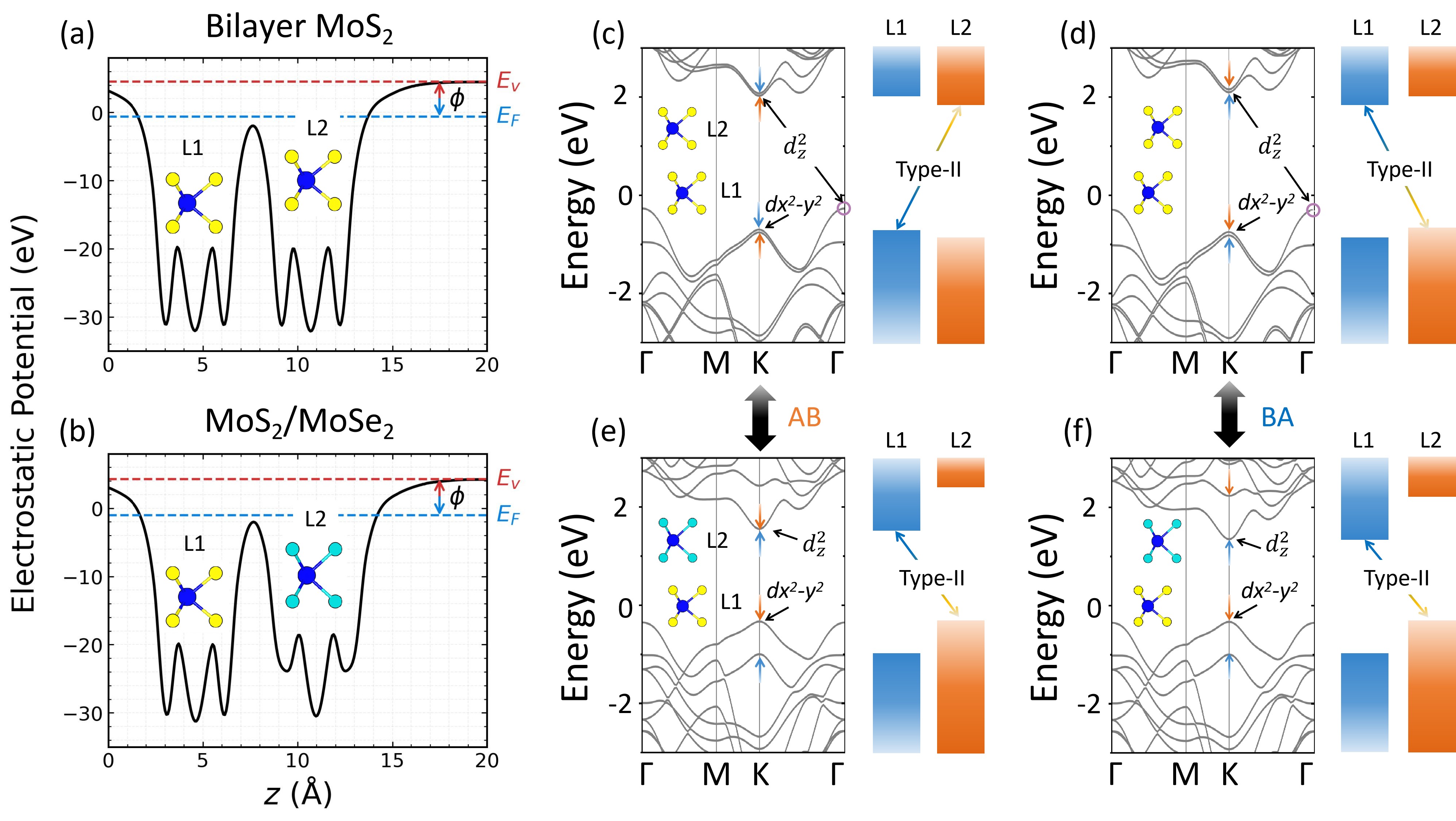}
\caption{Planar-averaged electrostatic potentials, quasiparticle band structures, and band alignments. (a)–(b) Out-of-plane electrostatic potential profiles for AB-stacked MoS$_2$ and MoS$_2$/MoSe$_2$, respectively; $E_v$ and $\phi$ indicate the vacuum level and work function. (c)–(f) Quasiparticle band structures calculated at the EVGW$_0$@PBE level for AB and BA stacking orders. The arrows highlight stacking-dependent shifts in the $d_{z^2}$ and $d_{x^2 - y^2}$ orbitals at the $K$ point. The right-hand schematics illustrate the resulting Type-II band alignments between layers L1 and L2.}
\label{fig:Fig2}
\end{figure*}

%--------------------------------------------Figure 2 ------------------------------------------------------
%--------------------------------------------Table ------------------------------------------------------
\begin{table*}[ht!]
\caption{Calculated vacuum levels ($E_\mathrm{vac}$), work functions ($\Phi$), electronic band gaps ($E_g$) obtained with PBE, PBE+SOC, HSE06, and EVGW$_0$@PBE, photoluminescence (PL) energies, and exciton binding energies ($E_b$) of the lowest excitonic transition for bilayer MoS$_2$ and MoS$_2$/MoSe$_2$ heterobilayers in AB and BA stackings.}
\centering
\small
\setlength{\tabcolsep}{4pt} % adjust column separation
\begin{tabular}{lcccccccccc}
\hline
System & Stacking & $E_\mathrm{vac}$ & $\Phi$ & $E_g^{\mathrm{PBE}}$ & $E_g^{\mathrm{PBE+SOC}}$ & $E_g^{\mathrm{HSE06}}$ & $E_g^{\mathrm{EVGW_0@PBE}}$ & PL Energy & $E_b$ \\
 & & (eV) & (eV) & (eV) & (eV) & (eV) & (eV) & (eV) & (eV) \\
\hline
Bilayer MoS$_2$   & AB & 4.49  & 5.10 & 1.23 & 1.22 & 1.74 & 2.26 & 2.08 & 0.18 \\
                  & BA & 4.49  & 5.10 & 1.25 & 1.24 & 1.77 & 2.29 & 2.07 & 0.22 \\
MoS$_2$/MoSe$_2$  & AB & 4.96  & 5.30 & 0.85 & 0.76 & 1.24 & 1.89 & 1.45 & 0.44 \\
                  & BA & 4.99  & 5.30 & 0.75 & 0.66 & 1.13 & 1.78 & 1.34 & 0.44 \\
\hline
\end{tabular}
\label{tab:workfunction}
\end{table*}
%--------------------------------------------Table -----------------------------------------------------------------
The electrostatic consequences of charge redistribution are quantified by the planar-averaged electrostatic potential $V(z)$, which is coupled to the charge density difference $\Delta \rho(z)$ via Poisson’s equation:
\begin{equation}
\frac{d^2 V(z)}{dz^2} = -\frac{\Delta \rho(z)}{\varepsilon_0}.
\end{equation}
The resulting out-of-plane electric field, $E_z = -dV/dz$, is determined by the first moment of the redistributed charge. A macroscopic potential shift $\Delta V$ across the interlayer distance $d_{\mathrm{int}}$ generates an effective internal field,
\begin{equation}
E_z \approx \frac{\Delta V}{d_{\mathrm{int}}}.
\end{equation}

For the AB-stacked $\mathrm{MoS}_2$ homobilayer (Fig.~\ref{fig:Fig2}a), the planar-averaged potential reveals a distinct layer-to-layer offset of $\sim 55~\mathrm{meV}$. To ensure methodological consistency, we applied a macroscopic averaging technique to filter out periodic variations; this confirms the offset is robust and yields an internal electric field of $\sim$ $0.086~\mathrm{V/nm}$ directed from L2 to L1. Inverting the registry to the BA configuration (Suppl. Mat., Fig.~S3a) reverses the sign of the potential difference ($-55~\mathrm{meV}$) and the vector of the internal field. Since the constituent layers are chemically identical, this polarization is an emergent phenomenon driven by registry-induced symmetry breaking and interlayer orbital hybridization.

In contrast, the $\mathrm{MoS}_2/\mathrm{MoSe}_2$ heterobilayer (Fig.~\ref{fig:Fig2}b) exhibits a qualitatively distinct electrostatic profile. The planar-averaged potential is dominated by large variations of the peaks ($\sim 24$ to $31~\mathrm{eV}$) corresponding to the ionic cores of the Se and Mo planes. While these features reflect the intrinsic electronegativity mismatch between S and Se, they represent local atomic-scale variations that do not contribute to a macroscopic dipole. Using the same macroscopic averaging technique applied to the homobilayer, the potential offset for the heterobilayer is found to be negligible (Suppl. Mat., Fig.~S3b), such that:
\begin{equation}
\Delta V_{\mathrm{hetero}} \approx 0, \quad E_z \approx 0.
\end{equation}

Consequently, despite the pronounced local chemical asymmetry, the $\mathrm{MoS}_2/\mathrm{MoSe}_2$ interface does not support a stacking-dependent, switchable internal electric field. The intrinsic chemical potential gradient effectively pins the electronic state, suppressing the collective, registry-dependent charge transfer required for the bistable dipolar states observed in the homobilayer.

\noindent
The calculated vacuum levels and work functions provide a complementary perspective on the electrostatic environment of these layered systems (Table~\ref{tab:workfunction}). Bilayer MoS$_2$ exhibits identical vacuum levels of 4.49~eV for both AB and BA stackings, with corresponding work functions of approximately 5.10~eV, consistent with the presence of a small, reversible interlayer dipole. In contrast, the MoS$_2$/MoSe$_2$ heterobilayer shows elevated vacuum levels of 4.96~eV (AB) and 4.99~eV (BA), and work functions of $\sim$5.3~eV, reflecting the chemical asymmetry of the interface. These values corroborate the electrostatic analysis since the homobilayer supports a bistable, stacking-dependent dipole manifested as the 55~meV potential step, whereas the heterobilayer lacks a switchable internal field and the macroscopic potential is dominated by intrinsic chemical offsets rather than stacking-induced charge redistribution.
% --------------------------------------------------------------------------------------------------
\subsubsection{Stacking-Dependent Electronic Structure and Band Alignment}

To establish the baseline electronic landscape of the isolated components, the fundamental band gaps, absolute vacuum-referenced valence-band maxima ($E_{\text{VBM}}$), and conduction-band minima ($E_{\text{CBM}}$) for both unstrained and strained monolayers of MoS$_2$ and MoSe$_2$ were evaluated (Table~S2). As summarized in Table~S2, mechanical strain induces a transition from direct to indirect band gaps and shifts the band edges relative to the vacuum level. Furthermore, the corresponding directional variations in electron and hole effective masses ($m_e^*$ and $m_h^*$ ) along the high-symmetry $K \to \Gamma$ and $K \to M$ paths (Table~S3) reveal that lattice deformation significantly modifies band curvature, directly impacting carrier mobility and effective mass anisotropy.

The electronic properties of homobilayer $\mathrm{MoS}_2$ and $\mathrm{MoS}_2/\mathrm{MoSe}_2$ heterobilayers were evaluated using PBE, PBE+SOC, and HSE06 functionals to investigate the impact of stacking configuration on the band gap $E_g$ and band-edge localization. To accurately account for many-body effects and the reduced dielectric screening characteristic of two-dimensional systems, eigenvalue-self-consistent quasiparticle calculations were performed within the $\mathrm{EV}GW_0@\mathrm{PBE}$ framework. Summations over unoccupied states were evaluated using 64 empty conduction bands, ensuring stable convergence of the polarizability matrix. Furthermore, while SOC is explicitly included at the PBE level (PBE+SOC), HSE06 and $GW_0$+BSE calculations were performed without SOC due to computational scaling limits. Because intra-atomic heavy-metal $d$-orbital SOC shifts are nearly invariant under rigid lateral sliding (AB vs. BA registries), leaving local intralayer atomic coordination and bonding geometry undisturbed, absolute relativistic contributions largely cancel out to first order when evaluating differential stacking-dependent energy shifts, ensuring that our comparative trends remain robust against this approximation, consistent with established first-principles benchmarks for transition metal dichalcogenide heterostructures \cite{Qiu2016,Gillen2018,Li2024}. The resulting band gap renormalization and the complex interplay between stacking registry and electronic structure are illustrated in Fig.~\ref{fig:Fig2}c--f, while a comprehensive comparison across the PBE, PBE+SOC, and HSE06 levels including fully layer- and orbital-resolved band structures is provided in the Suppl. Mat. (Figs.~S4--S7).

For the $\mathrm{MoS}_2$ homobilayer, both AB and BA stackings exhibit an indirect $\Gamma$--$K$ band gap. At the PBE level, the band gaps are approximately $E_g \approx 1.23$~eV for AB and $1.25$~eV for BA. Including SOC reduces the gaps slightly to about $1.22$~eV for AB and $1.24$~eV for BA, which is consistent with the moderate spin splitting of the Mo $d$ states. The HSE06 hybrid functional increases the gaps by roughly 40\%, yielding $E_g \approx 1.71$~eV for AB and $1.74$~eV for BA, while $\mathrm{EV}GW_0@\mathrm{PBE}$ further enhances them to $2.26$~eV for AB and $2.29$~eV for BA (Table~\ref{tab:workfunction}).

This indirect character originates from orbital-selective interlayer hybridization. The VBM and CBM at the $K$-point are dominated by in-plane Mo $d$ orbitals, which suppress interlayer coupling and effectively pin the localized states at $K$. Stacking-dependent shifts of these $d_{z^2}$ and $d_{x^2-y^2}$ states, $\Delta_v$ and $\Delta_c$, are highlighted by the arrows in Fig.~\ref{fig:Fig2}c,d, demonstrating how registry modulates local orbital energies.
In the AB configuration, the planar-averaged electrostatic potential profile (Fig.~\ref{fig:Fig2}a) reveals a distinct intrinsic potential offset of $\sim 55\,\mathrm{meV}$ between the layers, indicating a stacking-dependent dipole. This macroscopic electrostatic environment acts as a perturbation that modifies the electronic states, resulting in band offsets of $\Delta_v \approx 50\,\mathrm{meV}$ and $\Delta_c \approx 30\,\mathrm{meV}$ (as derived from the electronic band structure). This relationship is precisely inverted in the BA configuration due to the spatial reversal of the layer symmetry.

The $\Gamma$-point valence states are primarily composed of out-of-plane Mo $d_{z^2}$ and S $p_z$ orbitals, where the interlayer coupling produces a pronounced bonding--antibonding splitting of $\Delta_\Gamma=0.72$~eV (cf. Fig.~S5). 
The electronic coupling at $\Gamma$ can be estimated using a two-level approximation at the $\Gamma$-point. For a homobilayer the uncoupled VBM orbitals are degenerate. However, due to the static electric field there is an offset of $\Delta=0.05$ eV as given in the previous subsection. Thus the effective splitting is given by 
$\Delta_\Gamma = \sqrt{\Delta^2 + 4 t_\Gamma^2}$. From this expression the coupling (hopping matrix) follows as $t_\Gamma=0.36$~eV. 

The considerable magnitude of this extracted hopping parameter highlights the strong interlayer electronic hybridization occurring despite the predominantly vdW nature of the interlayer assembly. This coupling arises from the non-negligible spatial tails of out-of-plane atomic orbitals into the van der Waals gap—which is distinct from covalent bond formation—and this electronic overlap drives the large bonding--antibonding splitting ($\Delta_\Gamma$). This pushes the upper hybridized state above the localized uncoupled states at $K$, thereby establishing the definitive indirect band gap character of the homobilayer. Comparing the AB and BA configurations, the negligible variation observed in $\Delta_\Gamma$ demonstrates that while the local atomic registry directly modulates the layer-dependent potential offset, the interlayer hopping matrix element $t_\Gamma$ remains fundamentally constrained by the global equilibrium interlayer separation rather than the relative lateral atomic displacement.

As detailed in the quasiparticle band alignment scheme (Suppl. Mat., Fig.~S8), these registry-dependent potential offsets ultimately yield a type-II alignment that facilitates the formation of spatially separated interlayer excitons. These are denoted based on their specific interlayer character: $\mathrm{X}_{\mathrm{Ib}}$ (electron localized in the bottom layer), $\mathrm{X}_{\mathrm{It}}$ (electron localized in the top layer), and $\mathrm{X}_{\mathrm{IL}}$ (an exciton state associated with the global lattice registry).

In contrast, the $\mathrm{MoS}_2/\mathrm{MoSe}_2$ heterobilayer exhibits a robust direct type-II band alignment across all staggered registries. The potential profiles in Fig.~\ref{fig:Fig2}b show large atomic-scale oscillations; while such fluctuations are intrinsic to all van der Waals heterojunctions due to their periodic atomic structure, here they are clearly reflecting the chemical asymmetry between the layers. As depicted in the quasiparticle band structures (Fig.~\ref{fig:Fig2}e,f), where L1 ($\mathrm{MoS}_2$, blue) and L2 ($\mathrm{MoSe}_2$, orange) are color-coded, the CBM is strictly localized on L1, while the VBM resides on L2. This chemical potential mismatch between S and Se effectively overrides the registry-dependent potential shifts, preserving the direct $K$--$K$ gap and ensuring that the intrinsic band offsets indicated by the work functions ($\phi$) and vacuum levels ($E_v$) in Fig.~\ref{fig:Fig2}b remain the primary drivers of electronic localization.

% --------------------------------------------------------------------------------------------------
\subsubsection{Optical Properties and Excitonic Response}
% --------------------------------------------------------------------------------------------------

\begin{figure*}[ht!]
\centering
\includegraphics[width=0.7\textwidth]{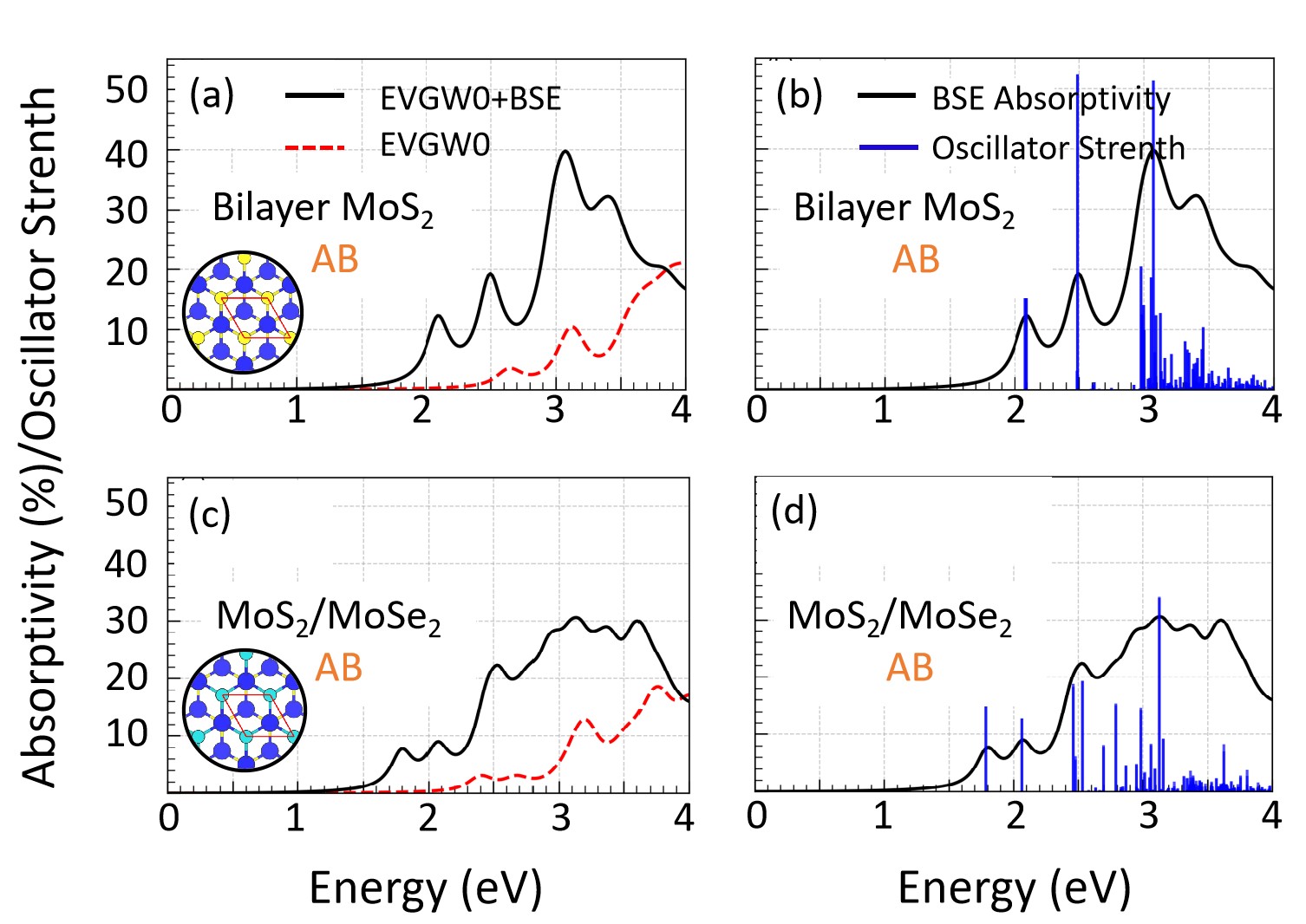}
\caption{
Many-body excitonic absorption spectra and stacking-dependent photoluminescence (PL).(a, c) Absorptivity of AB-stacked bilayer MoS$_2$ and MoS$_2$/MoSe$_2$ calculated at the EVGW$_0$ (red dashed) and EVGW$_0$+BSE (black solid) levels, illustrating the significant excitonic enhancement and band shifts. (b, d) BSE absorptivity (black, broadened with gaussian of width 0.1 eV) with individual transitions and oscillator strengths (blue bars).}
\label{fig:absorption}
\end{figure*}

Excitonic effects were evaluated by solving the Bethe--Salpeter equation (BSE) atop the quasiparticle (QP) landscape (Fig.~\ref{fig:absorption}). In these reduced-dimensional systems, suppressed dielectric screening enhances the electron--hole ($e\text{--}h$) interaction, yielding sizable exciton binding energies, $E_b^{\mathrm{exc}} = E_g^{\mathrm{QP}} - E_{\mathrm{opt}}$, and a marked renormalization of the absorption onset. The dimensionless oscillator strength $f_{\mathrm{exc}}$, which quantifies transition probability and spectral weight, is defined as
\begin{equation}
f_{\mathrm{exc}} = \frac{2m_e\omega_{\mathrm{exc}}}{\hbar} \big| \langle \psi_c | \hat{\mathbf{e}} \cdot \mathbf{r} | \psi_v \rangle \big|^2,
\end{equation}
where $m_e$ is the electron mass, $\hbar$ is the reduced Planck constant, and $\omega_{\mathrm{exc}}$ is the excitonic transition frequency. The energy-dependent prefactor ensures proper dimensional cancellation of the position matrix element, yielding the dimensionless values reported from our BSE calculations.

To obtain a box-independent optical response in 2D slab calculations, the spectra are evaluated in terms of the 2D sheet absorptivity
\begin{equation}
A(\omega) \approx \frac{\omega L_z}{c} \operatorname{Im}\{\varepsilon(\omega)\},
\end{equation}
where $L_z$ is the out-of-plane lattice parameter of the periodic unit cell (combining the slab thickness and a $\sim 20\,\text{\AA}$ vacuum region along $z$ to suppress spurious interactions between periodic images). Multiplying $\operatorname{Im}\{\varepsilon(\omega)\}$ by each structure's respective out-of-plane cell height $L_z$ ($\sim 32\text{--}37\,\text{\AA}$ across the optimized bilayers and trilayers, Table~S1) cancels the artificial $1/L_z$ scaling of the 3D macroscopic dielectric function and ensures consistent intensity comparisons across all stackings. In addition, the BSE matrix construction inherits quasiparticle inputs evaluated with 64 empty conduction bands to ensure numerical stability of the excitonic eigenvalues.

In the MoS$_2$ homobilayer, the optical response is dominated by intense intralayer excitons ($f_{\mathrm{exc}} \approx 19$) at $2.085\text{ eV}$ ($\text{AB}$) and $2.073\text{ eV}$ ($\text{BA}$), corresponding to exciton binding energies $E_b^{\mathrm{exc}} \approx 0.18\text{ eV}$ and $0.22\text{ eV}$, respectively (Table~\ref{tab:workfunction}; Fig.~\ref{fig:absorption}a,b and Fig.~S9(a,b)). Despite stacking-dependent band offsets ($\Delta_v$ and $\Delta_c$), the small $\sim 12\text{ meV}$ difference indicates that these intralayer states are largely decoupled from net interlayer polarization. This stability is consistent with the in-plane $d_{x^2-y^2}$ and $d_{xy}$ orbital character at the $K$ valley (Fig.~\ref{fig:Fig2}c,d), which preserves strong intralayer $e\text{--}h$ overlap and prevents significant interlayer leakage (Fig.~\ref{fig:absorption}e,f).

Conversely, the MoS$_2$/MoSe$_2$ heterobilayer exhibits strongly suppressed oscillator strength ($f_{\mathrm{exc}} \approx 0$) for the lowest-energy transitions (Fig.~\ref{fig:absorption}c,d). This optical darkness is a direct spectroscopic signature of the intrinsic type-II band alignment, where band offsets across the $\text{S}$--$\text{Se}$ interface spatially separate holes in MoSe$_2$ and electrons in MoS$_2$, reducing the transition dipole matrix element while yielding exciton binding energies up to $E_b^{\mathrm{exc}} \approx 0.44\text{ eV}$ (Table~\ref{tab:workfunction}). The resulting interlayer excitons ($X_{\mathrm{Ib}}$, $X_{\mathrm{It}}$, and $X_{\mathrm{IL}}$; Suppl.\ Mat., Fig.~S8) show a pronounced $\sim 107\text{ meV}$ shift across stackings—nearly an order of magnitude larger than the variation observed in the homobilayer.

At higher energies ($\sim 2.45\text{ eV}$), hybrid excitonic states emerge with finite oscillator strengths. The reduced hybrid intensity in the $\text{BA}$ stacking relative to $\text{AB}$ (cf. Fig.~\ref{fig:absorption}d and Suppl.\ Mat., Fig.~S9(c,d)) correlates with suppressed interlayer orbital hybridization ($t_{\Gamma}$) and modified charge redistribution in the CDD profiles (Fig.~\ref{fig:Fig1-1}). These results confirm that lateral sliding in TMD heterostructures provides an effective degree of freedom to modulate wavefunction delocalization across the van der Waals interface.
%--------------------------------------------Figure 3 ------------------------------------------------------
%--------------------------------------------Figure 3 ------------------------------------------------------

% --------------------------------------------------------------------------------------------------
\subsection{Trilayer 2L-MoS$_2$/MoSe$_2$ Heterostructures}
% --------------------------------------------------------------------------------------------------
\subsubsection{Stacking-Dependent Polarization in Trilayer 2L-MoS$_2$/MoSe$_2$ Heterostructures}

% Figure: Atomic configurations and CDD isosurfaces
% --------------------------------------------------------------------------------------------------
\begin{figure*}[ht!]
\centering
\includegraphics[width=1\textwidth]{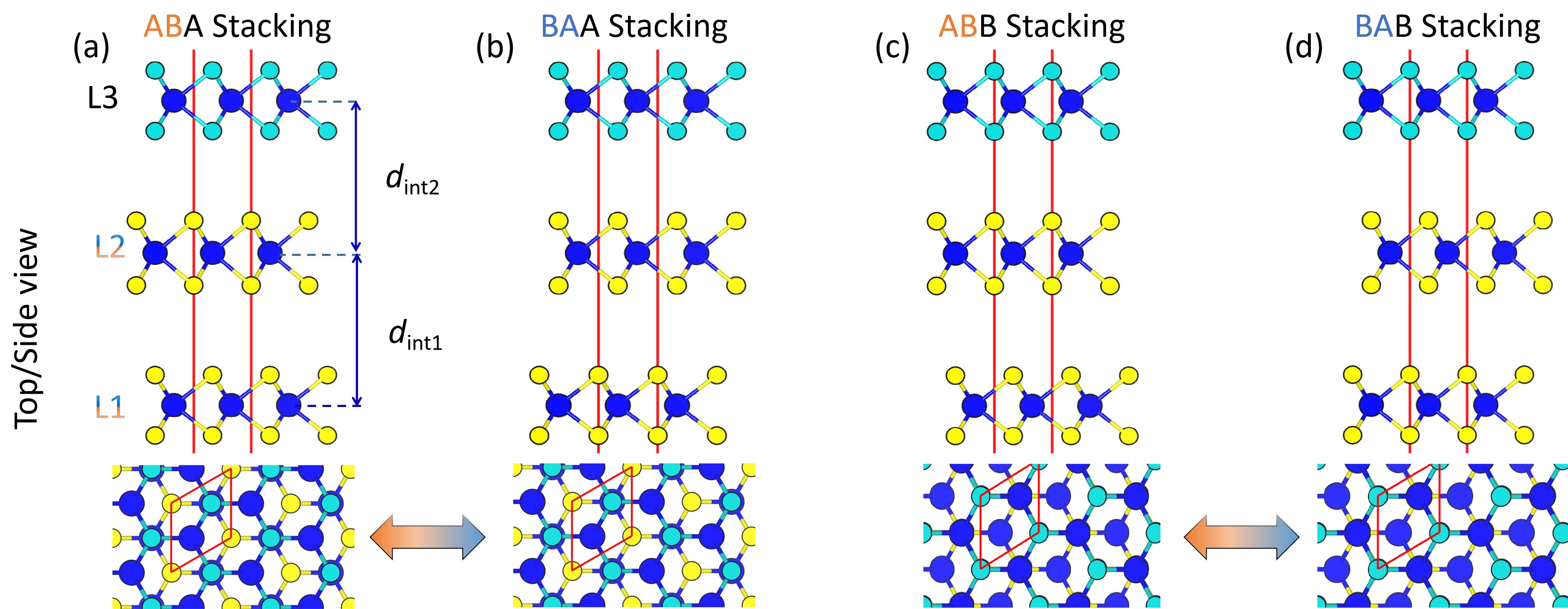}
\caption{
Atomic configurations of 2L-MoS$_2$/MoSe$_2$ heterostructures for different stacking sequences. Panels (a--d) show the ABA, BAA, ABB, and BAB configurations, respectively, in both top and side views. The interlayer distances $d_{\mathrm{int1}}$ and $d_{\mathrm{int2}}$ denote the separations between L1--L2 and L2--L3, highlighting the structural variation induced by stacking order.
}
\label{fig:stacking_structures}
\end{figure*}

The evolution of spontaneous out-of-plane polarization was investigated in trilayer heterostructures consisting of a $\text{MoSe}_2$ monolayer atop a $\text{MoS}_2$ homobilayer ($\text{L1-MoS}_2/\text{L2-MoS}_2/\text{L3-MoSe}_2$). This mixed-chalcogenide architecture introduces additional interlayer degrees of freedom, enabling precise modulation of the electric dipole through the interplay of multiple van der Waals interfaces. High-symmetry stacking sequences ($\text{ABA}$, $\text{BAA}$, $\text{ABB}$, and $\text{BAB}$) are analyzed in Fig.~\ref{fig:stacking_structures}a--d, while the asymmetric $\text{AAB}$ and $\text{BBA}$ configurations are discussed in the Supplementary Materials, Fig.~S11.

The in-plane lattice constant $a$ for all trilayer configurations converges within a narrow range of 3.202--3.208\,\AA, as summarized in Table~S1. This indicates that the observed variations in electronic and polarization properties are primarily driven by registry-dependent interlayer interactions rather than residual epitaxial strain. While the relaxed out-of-plane supercell dimension $c$ is approximately 32.19\,\AA{} for the configuration initialized from the $\text{AAA}$ registry (compared with 36.29--36.61\,\AA{} for the staggered configurations, utilizing a consistent vacuum region as noted in the Computational Details), the total lattice constant of a slab model includes the vacuum padding and is therefore not an appropriate physical measure of interlayer contraction on its own. Instead, structural compaction and interlayer proximity are rigorously evaluated via the actual interlayer distances, $d_{\mathrm{int1}}$ (between the two $\text{MoS}_2$ layers, $\text{L1}$ and $\text{L2}$) and $d_{\mathrm{int2}}$ (across the $\text{MoS}_2$--$\text{MoSe}_2$ heterointerface between $\text{L2}$ and $\text{L3}$).

Side views of the atomic arrangements (Fig.~\ref{fig:stacking_structures}a--d) further illustrate that $d_{\mathrm{int1}}$ and $d_{\mathrm{int2}}$ show strong sensitivity to stacking sequence. For instance, $d_{\mathrm{int1}}$ contracts to 6.41\,\AA{} in $\text{ABA}$ and 6.39\,\AA{} in $\text{BAA}$ relative to the $\text{AAA}$-initialized high-energy configuration (6.70\,\AA{}), highlighting how lateral shifts in staggered registries alleviate Pauli steric repulsion and allow adjacent layers to nest more closely. Conversely, $d_{\mathrm{int2}}$, which governs the heterofacial $\text{S}$--$\text{Se}$ interface, exhibits distinct values (e.g., 6.64\,\AA{} in $\text{ABA}$ versus 6.96\,\AA{} in the $\text{AAA}$-initialized state), reflecting the larger covalent radius of Selenium. This local structural relaxation enhances interlayer orbital hybridization, facilitates charge redistribution across the heterojunction, and ultimately modulates the net out-of-plane polarization. Table~S1 quantitatively captures these trends in $a$, $d_{\mathrm{int1}}$, and $d_{\mathrm{int2}}$ across all systems, providing a structural framework to rationalize the stacking-dependent electronic and polar properties.

\begin{figure*}[ht!]
\centering
\includegraphics[width=1\textwidth]{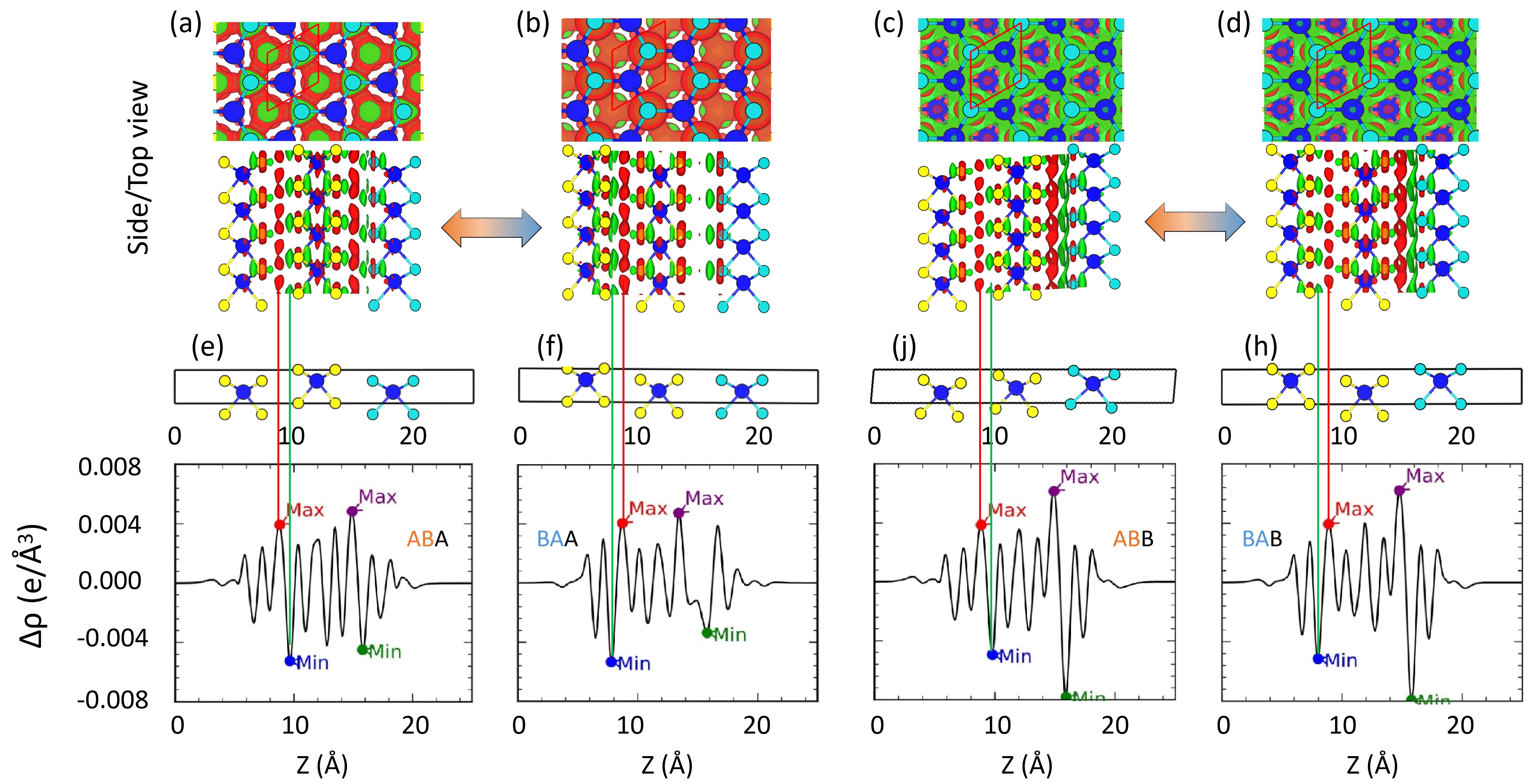}
\caption{
Charge redistribution in 2L-MoS$_2$/MoSe$_2$ heterostructures for different stacking sequences. Panels (a--d) show the charge density difference ($\Delta \rho$) isosurfaces for ABA, BAA, ABB, and BAB configurations, respectively, where yellow and cyan regions represent electron accumulation and depletion (isovalue: 0.001 e/\AA$^3$). Panels (e--h) present the corresponding planar-averaged $\Delta \rho(z)$ profiles across the trilayer, revealing the interfacial dipole characteristics associated with each stacking sequence.
}
\label{fig:charge_density}
\end{figure*}

The interlayer charge redistribution was quantified using the planar-averaged CDD
\begin{equation}
\Delta \rho(z) = \rho_{\rm{trilayer}}(z) - \sum_{i=1}^{3} \rho_{L_i}(z).
\end{equation}
The 3D isosurfaces (Fig.~\ref{fig:charge_density}a--d) and corresponding 1D profiles (Fig.~\ref{fig:charge_density}e--h) provide a clear spatial map of this redistribution. Analysis of the CDD profiles shows that the L2/L3 (MoS$_2$/MoSe$_2$) interface maintains a robust, stacking-insensitive maximum-to-minimum sequence. This indicates that the interfacial dipole at the heterojunction is primarily dictated by the intrinsic chemical potential mismatch between S and Se, effectively pinning the polarization regardless of the underlying bilayer registry.

In contrast, the L1/L2 (MoS$_2$/MoS$_2$) homobilayer interface acts as a tunable degree of freedom. In the $\text{AAA}$-initialized configuration, the L1/L2 junction exhibits a symmetric charge distribution with vanishing net dipole (Suppl. Mat., Fig.~S10). In $\text{ABA}$ and $\text{ABB}$ sequences, the L1/L2 interface adopts an AB-type registry, producing a maximum-to-minimum CDD sequence and a corresponding upward dipole vector (Fig.~\ref{fig:charge_density}e,g). Conversely, $\text{BAA}$ and $\text{BAB}$ stackings invert this sequence to minimum-to-maximum (Fig.~\ref{fig:charge_density}f,h), generating a downward dipole vector analogous to BA-type homobilayers.

The $\text{AAB}$ and $\text{BBA}$ configurations further illustrate the role of structural relaxation in dipole formation. In $\text{AAB}$, the initial AA-like registry of the L1/L2 interface is energetically unfavorable due to S--S chalcogen repulsion; structural relaxation induces lateral sliding of L1 toward a BA-like arrangement, breaking out-of-plane symmetry and producing a downward dipole. In $\text{BBA}$, the L1/L2 interface occupies a near-equilibrium staggered position, resulting in a symmetric extrema sequence and negligible net polarization (Suppl. Mat., Fig.~S11).

Collectively, these results demonstrate that the 2L-MoS$_2$/MoSe$_2$ trilayer behaves as a multistable dipolar system. By delineating the local dipole orientations at each interface—whether pointing from the MoS$_2$ layers toward MoSe$_2$ or in reverse—Fig.~\ref{fig:charge_density} illustrates how the internal electric field can be modulated across stacking sequences.

To distinguish energetically favorable relaxed stacking configurations from higher-energy arrangements, full structural relaxations were performed for the considered stacking sequences. The structure initialized in the high-symmetry $\text{AAA}$ registry undergoes interlayer sliding during relaxation and converges to a relatively high-energy configuration ($\Delta E = 143.10$ meV relative to $\text{BAB}$; Table~S4), suggesting unfavorable interlayer interactions in this registry. In contrast, the $\text{ABA}$, $\text{BAA}$, $\text{ABB}$, and $\text{BAB}$ configurations converge to distinct relaxed structures with substantially lower energies. A quantitative comparison of their total energies (Table~S4) reveals that the $\text{BAB}$ and $\text{ABB}$ configurations differ by only $0.01$ meV/cell, which is significantly smaller than the estimated numerical uncertainty associated with $k$-point convergence ($\sim 0.3$ meV/cell). These two configurations are therefore considered energetically degenerate within the numerical accuracy of our DFT calculations, and no definitive energetic ordering is assigned. The $\text{ABA}$ and $\text{BAA}$ configurations lie $4.70$ and $75.44$ meV/cell, respectively, above the $\text{BAB}$ reference. The distinct relaxed interlayer distances ($d_{\mathrm{int1}}$ and $d_{\mathrm{int2}}$) and charge redistribution profiles further highlight the strong dependence of the local structural and dipolar characteristics on stacking registry. The relationship between stacking registry, symmetry breaking, and polarization in layered van der Waals materials has been discussed theoretically, including in the context of stacking-induced ferroelectricity and multiple polarization states in multilayer systems \cite{meng2022_7696, ji2023_146801}. Importantly, we emphasize that the identification of distinct relaxed configurations should not be interpreted as direct evidence of switchable multistability or kinetic stability against interlayer sliding. Establishing these properties requires explicit evaluation of the interlayer sliding energy landscape, minimum-energy transition pathways, and associated switching barriers, which are beyond the scope of the present static DFT investigation. Previous experimental observations of stacking-dependent domains and sliding ferroelectricity in transition-metal dichalcogenide heterostructures \cite{Li2021} provide relevant physical context for the potential importance of these stacking-dependent properties. Accordingly, our present results establish the influence of local stacking registry on structural relaxation, electronic band alignment, charge redistribution, and excitonic properties, while the potential implications for polarization switching and non-volatile memory remain prospective and require further investigation.
% --------------------------------------------------------------------------------------------------
%--------------------------------------------------------------------------------------------------
\subsubsection{Quasiparticle Band Structure Modulation by Stacking in Trilayer 2L-MoS$_2$/MoSe$_2$}
%--------------------------------------------------------------------------------------------------
\begin{figure}[t]
\centering
\includegraphics[width=0.8\linewidth]{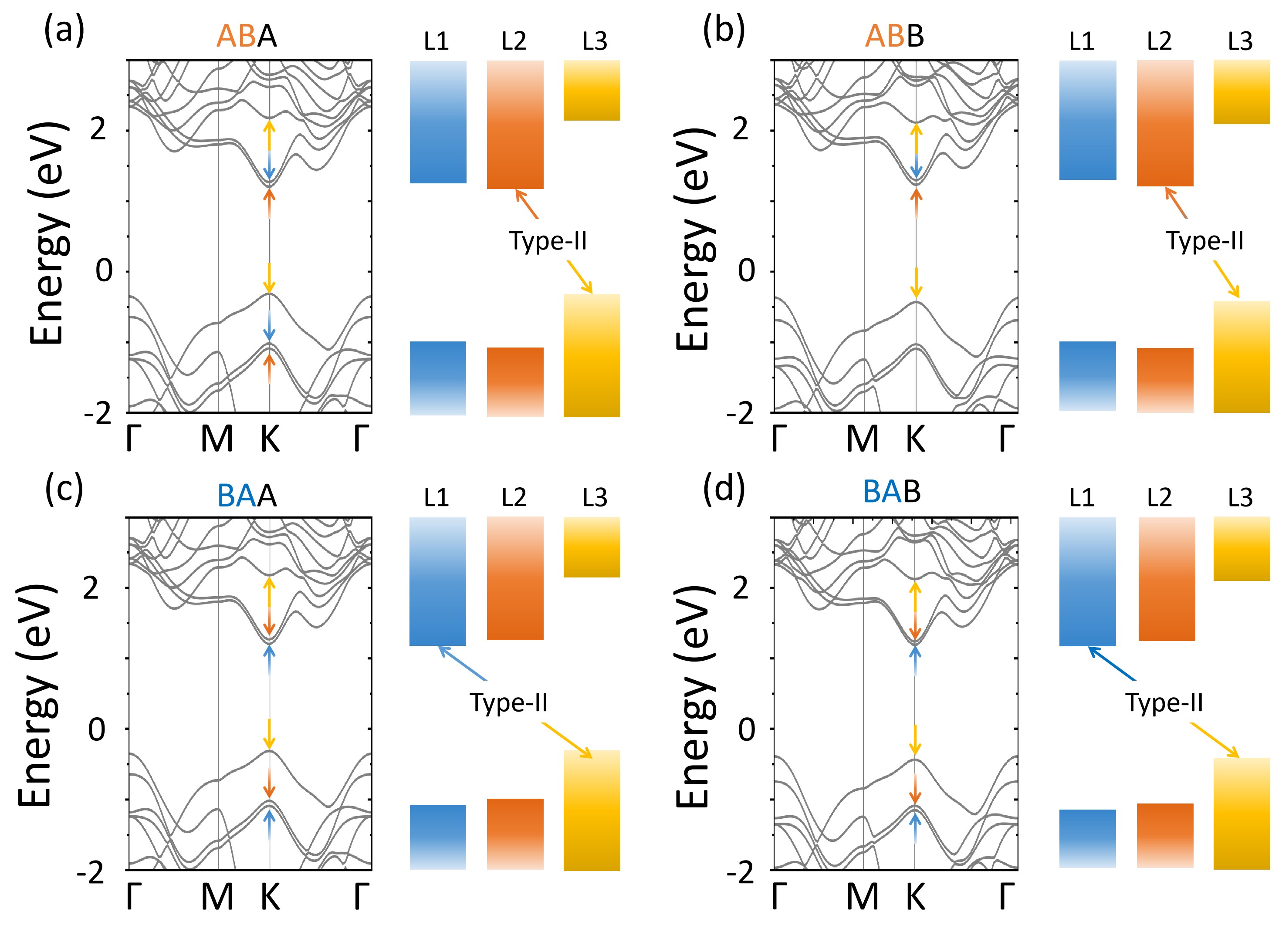}
\caption{Quasiparticle electronic structures and band alignments of trilayer 2L-MoS$_2$/MoSe$_2$ heterostructures. (a)--(d) EVGW$_0$ quasiparticle band structures and corresponding type-II band alignment schematics for ABA, ABB, BAA, and BAB stacking sequences. The arrows at the $K$ point highlight the stacking-dependent localization of the conduction-band minimum (CBM) and valence-band maximum (VBM) across the L1, L2, and L3 layers.}
\label{fig:2LMoS2-MoSe2-band}
\end{figure}

The electronic properties of 2L-MoS$_2$/MoSe$_2$ trilayer heterostructures show a clear and systematic sensitivity to atomic registry, as summarized in Table~\ref{tab:bandgaps_PL_2LMoS2_MoSe2}, Fig.~\ref{fig:2LMoS2-MoSe2-band}, and Figs.~S12--S14. At the PBE+SOC level, the fundamental band gaps span 0.68--0.83~eV, corresponding to a $\sim$22\% modulation driven by stacking-dependent interlayer coupling. This persistence across electronic structure levels indicates that atomic registry acts as an effective internal boundary condition for the electronic landscape of these quasi-2D heterostructures.

A systematic comparison across computational levels shows consistent band-gap scaling. Moving from PBE to HSE06 increases the band gap by $\sim$60--70\%, while many-body quasiparticle corrections at the EVGW$_0$@PBE level add a further $\sim$20--25\% increase relative to HSE06. The resulting quasiparticle gaps (1.51--1.60~eV) exceed PBE+SOC values by roughly 93--120\%, reflecting the combined impact of exact exchange and reduced dielectric screening in two dimensions.

\begin{figure}[t]
\centering
\includegraphics[width=1\linewidth]{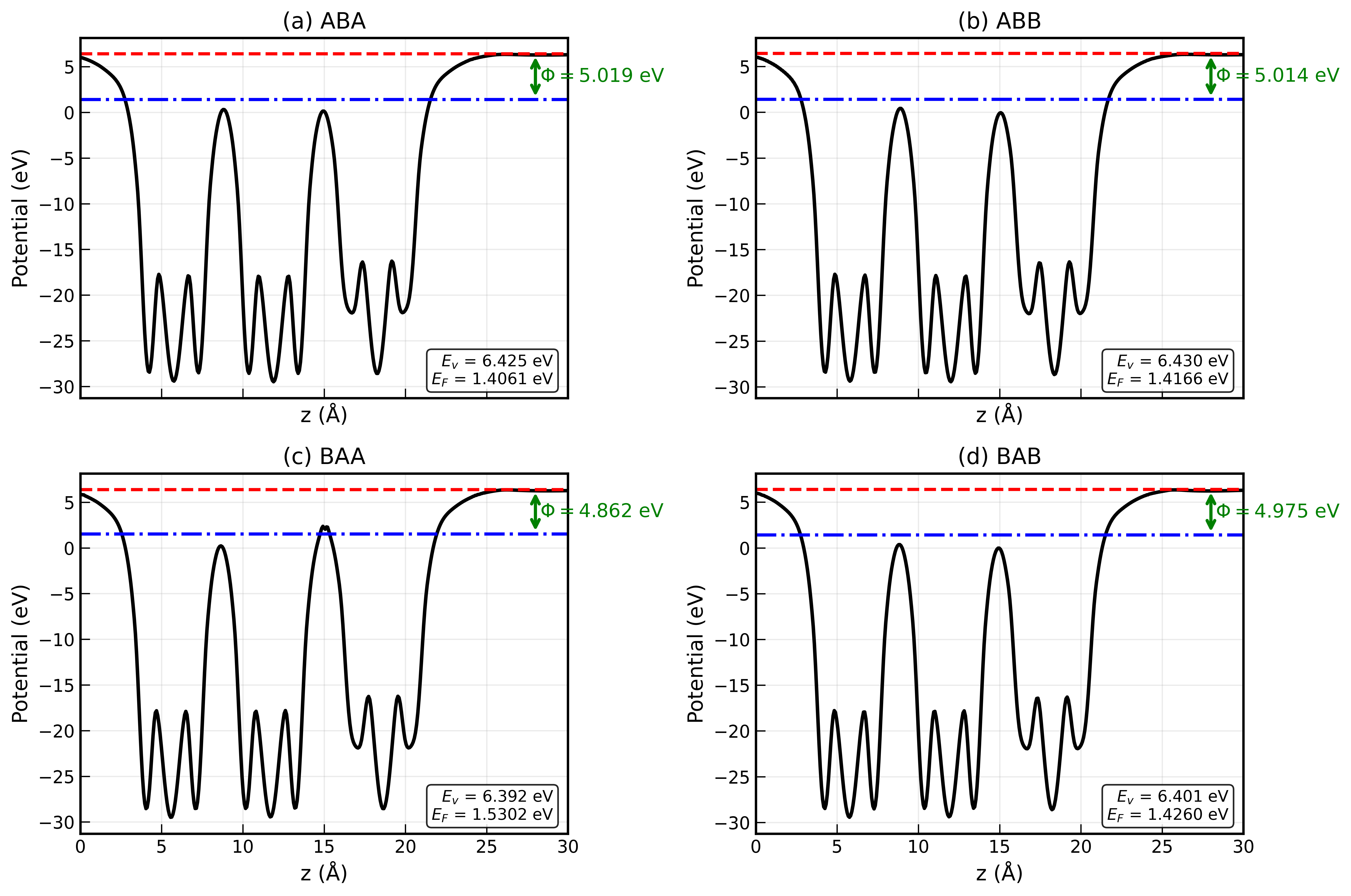}
\caption{Planar-averaged electrostatic potential profiles $V(z)$ and work function analysis for trilayer 2L-MoS$_2$/MoSe$_2$ heterostructures. (a)--(d) Potential energy landscapes along the $c$-axis for ABA, ABB, BAA, and BAB stackings. Red dashed lines denote the vacuum level ($E_{\mathrm{v}}$), blue dash-dotted lines indicate the Fermi level ($E_{\rm F}$), and green bidirectional arrows represent the stacking-dependent work function ($\Phi$). The inset information boxes provide the corresponding energetic references obtained after applying a macroscopic averaging filter.}
\label{fig:potential_profiles}
\end{figure}

The stacking-dependent electronic structure is illustrated in Fig.~\ref{fig:2LMoS2-MoSe2-band}(a--d), which displays EVGW$_0$ band dispersions and type-II alignment schematics for $\text{ABA}$, $\text{ABB}$, $\text{BAA}$, and $\text{BAB}$. Complementary planar-averaged electrostatic potentials $V(z)$ (Fig.~\ref{fig:potential_profiles}(a--d)) show that atomic registry introduces a configuration-dependent internal electrostatic potential profile, reflected in systematic shifts of the vacuum level $E_{\mathrm{vac}}$ relative to the Fermi level $E_{\rm F}$. These shifts correspond to work-function ($\Phi$) values of 5.019~eV ($\text{ABA}$), 5.014~eV ($\text{ABB}$), 4.862~eV ($\text{BAA}$), and 4.975~eV ($\text{BAB}$), showing that local electrostatics varies systematically with stacking order.

This registry-dependent electrostatic landscape stems from stacking-dependent structural relaxation at the $\text{MoS}_2$/$\text{MoS}_2$ interface, where Pauli repulsion and van der Waals attraction balance differently across stackings. The resulting charge redistribution modifies $V(z)$ and generates modest internal potential variations relative to $E_{\rm F}$. Rather than acting as a uniform rigid shift of all bands, this internal potential bias competes with interlayer orbital hybridization within the $\text{MoS}_2$ bilayer, while the $\text{MoS}_2$/$\text{MoSe}_2$ interface remains more rigid and anchors the type-II heterojunction alignment.

This interplay is most apparent in the conduction sector. Layer- and momentum-resolved projections at the $K$ point show that the VBM remains localized on the $\text{MoSe}_2$ layer across the examined stackings, providing a stable valence reference for the type-II alignment. In contrast, the CBM lies within a near-degenerate manifold spanning the $\text{MoS}_2$ bilayer, making its internal splitting and layer weight sensitive to interlayer coupling and local potential asymmetry.

Specifically, the interlayer conduction-band splitting decreases from 62.7--65.8~meV ($\text{ABA}$/$\text{ABB}$) to 53.4--48.5~meV ($\text{BAA}$/$\text{BAB}$), reflecting weaker internal splitting across the $\text{MoS}_2$ bilayer. While the mean/envelope energy of the conduction-band manifold shifts only modestly across stackings, the internal splitting and layer projection between the $\text{MoS}_2$ layers redistribute within this envelope.

This redistribution follows the same electrostatic trends seen in $V(z)$, where changes in $E_{\mathrm{vac}} - E_{\rm F}$ track the work-function shifts. The resulting local potential bias tunes the balance between interlayer hybridization and layer polarization across the conduction manifold without disrupting the $\text{MoSe}_2$-localized VBM.

Spin--orbit coupling further refines the band-edge structure at the PBE level by splitting valence-band states at the $K$ point and tuning direct-gap character across stackings. Across all computational levels, the quasiparticle gaps and optical transition energies align with the stacking-dependent excitonic trends summarized in Table~\ref{tab:bandgaps_PL_2LMoS2_MoSe2}, with $\text{ABB}$ exhibiting the largest EVGW$_0$ gap ($1.60\text{ eV}$).

Overall, stacking registry in 2L-MoS$_2$/MoSe$_2$ trilayers acts as an internal electrostatic boundary condition linking atomic arrangement to local dipolar fields and band-edge orbital character. This interplay pairs a relatively steady valence reference on $\text{MoSe}_2$ with a registry-tunable conduction manifold on the $\text{MoS}_2$ bilayer, providing a practical route to tailor internal potentials, band-edge character, and optical response in layered van der Waals heterostructures.

%--------------------------------------------------------------------------------------------------------------------------

%The stacking-dependent modulation originates from the combined effects of interlayer orbital hybridization and local electrostatic potentials. To quantify these effects, layer-resolved potential offsets $\Delta V_{ij}$ ($i,j \in \{1,2,3\}$) are extracted by applying a macroscopic average filter 
%(with a window width equal to the $c$-axis lattice parameter) 

% --------------------------------------------------------------------------------------------------
\begin{table*}[ht]
\centering
\caption{
Calculated band gaps ($E_g$, in eV) using PBE, PBE+SOC, HSE06, and EVGW$_0$ ($E_g^{\text{QP}}$), optical/PL transition energies ($E_{\text{opt}}$), exciton binding energies ($E_b$), along with gap type and photoluminescence (PL) energies for different stacking configurations of 2L-MoS$_2$/MoSe$_2$ trilayer heterostructures. Values of $E_b$ demonstrate systematic stacking-dependent modulation across registries, reflecting subtle registry-induced variations in interlayer distances and local dielectric screening. 
}
\label{tab:bandgaps_PL_2LMoS2_MoSe2}

\resizebox{\textwidth}{!}{
\begin{tabular}{lcccccccc}
\hline
\textbf{Stacking} & \textbf{PBE} & \textbf{PBE+SOC} & \textbf{HSE06} & \textbf{EVGW$_0$ ($E_g^{\text{QP}}$)} & \textbf{PL / $E_{\text{opt}}$ (eV)} & \textbf{$E_b$ (eV)} & \textbf{Gap Type} \\
\hline
AAA & 0.82 & 0.74 & 1.25 & 1.55 & 1.362 & 0.188 & Direct \\
ABA & 0.82 & 0.73 & 1.21 & 1.51 & 1.287 & 0.223 & Direct \\
ABB & 0.86 & 0.83 & 1.31 & 1.60 & 1.391 & 0.209 & Indirect (PBE)/Direct (SOC, HSE06) \\
BAA & 0.77 & 0.68 & 1.19 & 1.51 & 1.356 & 0.154 & Direct \\
BAB & 0.82 & 0.78 & 1.28 & 1.58 & 1.395 & 0.185 & Indirect (PBE)/Direct (SOC, HSE06) \\
AAB & 0.83 & 0.79 & 1.28 & 1.59 & 1.398 & 0.192 & Indirect (PBE)/Direct (SOC, HSE06) \\
BBA & 0.82 & 0.73 & 1.22 & 1.52 & 1.278 & 0.242 & Direct \\
\hline
\end{tabular}
}
\end{table*}
%-----------------------------------------------------------------------------------------------

\subsubsection{Optical Properties and Excitonic Response}
%--------------------------------------------------------------------------------------------------

\begin{figure}[t]
\centering
\includegraphics[width=0.6\linewidth]{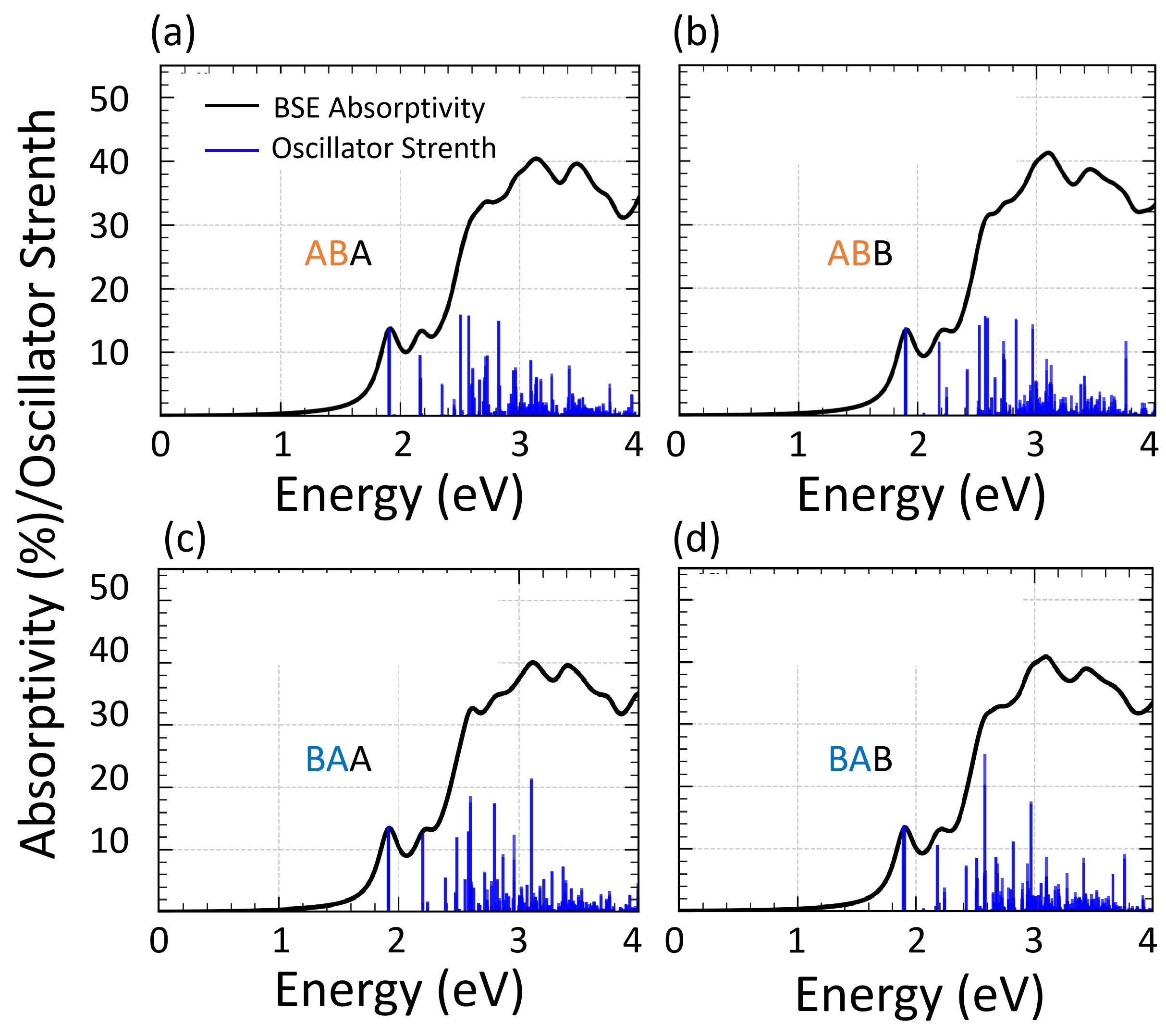}
\caption{Excitonic absorption of trilayer 2L-MoS$_2$/MoSe$_2$ heterostructures. (e)--(h) Many-body excitonic absorption spectra (black solid lines, Gaussian broadening with width 0.1 eV) and oscillator strengths (blue bars) calculated at the EVGW$_0$+BSE level. The variations in the low-energy excitonic peaks reflect the modulation of interlayer coupling and spatial charge separation induced by the different stacking registries.}
\label{fig:optical-2L}
\end{figure}

The optical response of the 2L-MoS$_2$/MoSe$_2$ trilayer was systematically evaluated to elucidate the impact of stacking-dependent polarization on the many-body excitonic landscape. The specific electronic structure is reflected in the many-body excitonic absorption spectra (Fig.~\ref{fig:optical-2L}a--d), calculated at the EVGW$_0$+BSE level using a Gaussian broadening of $\rm{0.1}$~eV. The black solid lines represent the absorption spectra, while the blue bars denote the oscillator strengths ($f_{\rm{exc}}$); notably, the absorptivity is the result of the convolution of these oscillator strengths with the broadening function. For all configurations, the lowest-energy peaks exhibit strongly suppressed oscillator strengths, a hallmark of spatially separated interlayer excitons. The variation in intensity across different stackings reflects differences in interlayer coupling; for example, ABB and BAB configurations exhibit higher oscillator strengths than BAA, indicating enhanced wavefunction overlap across the vdW gap.

To explicitly address the influence of stacking modes on excitonic properties, exciton binding energies ($E_b$) were quantitatively determined across all configurations as the difference between the quasiparticle band gap and the optical transition energy ($E_b = E_g^{\text{QP}} - E_{\text{opt}}$). As summarized in Table~\ref{tab:bandgaps_PL_2LMoS2_MoSe2}, $E_b$ exhibits a clear stacking-dependent modulation, ranging from 0.154~eV (BAA) to 0.242~eV (BBA). This variation is governed by registry-induced changes in interlayer distance and local orbital overlap, which subtly modulate the electron--hole Coulomb interaction while maintaining consistent macroscopic dielectric screening across the van der Waals stack.

Table~\ref{tab:bandgaps_PL_2LMoS2_MoSe2} summarizes the PL energies for the lowest-energy transitions, which span a wide spectral range from 1.278 to 1.398~eV. These variations originate from the interplay between the intrinsic type-II band alignment and stacking-induced electrostatic potential modulations at the MoS$_2$ bilayer interfaces (Suppl.\ Mat., Fig.~S8). In these heterostructures, the lowest-energy excitations are dominated by interlayer excitons (ILXs), characterized by spatial separation of charge carriers, with holes localized in the MoSe$_2$ valence band and electrons residing in the MoS$_2$ bilayer. Consequently, the transition energies are highly sensitive to internal electric fields arising from stacking-dependent dipole moments. This can already be seen from the EVGW$_0$+BSE absorption spectra (Fig.~\ref{fig:optical-2L}a--d) where AB-type stackings (ABA and ABB) exhibit stronger excitonic peaks compared to BA-type configurations (BAA and BAB), reflecting enhanced electron--hole wavefunction overlap. To quantitatively assess the effect of polarization reversal, the average ILX energies for the two stacking families were evaluated. This approximately accounts for variable stacking configurations between MoS$_2$ and MoSe$_2$ due to lattice mismatch and domain disorder in actual experiments. For AB-type configurations, the characteristic energy is defined as $E_{\rm{AB}}^{\rm{avg}} = (E_{\rm{ABA}} + E_{\rm{ABB}})/2 \approx 1.339~\rm{eV}$, whereas for BA-type configurations it is given by $E_{\rm{BA}}^{\rm{avg}} = (E_{\rm{BAA}} + E_{\rm{BAB}})/2 \approx 1.375~\rm{eV}$. This results in an interlayer excitonic energy shift $\Delta E_{\rm{ILX}} = E_{\rm{BA}}^{\rm{avg}} - E_{\rm{AB}}^{\rm{avg}} \approx 36~\rm{meV}$.

\begin{figure}[t]
\centering
\begin{tikzpicture}[scale=1.1, >=stealth]

% --- STYLING ---
\tikzstyle{level} = [ultra thick, line width=2pt]
\tikzstyle{arrow} = [->, thick, green!90!black]
\tikzstyle{labelnode} = [font=\sffamily\small, align=center]
\tikzstyle{dim} = [<->, thick, black!90]

% ==============================================================================
% --- AB-TYPE (LEFT) ---
% ==============================================================================
\begin{scope}[xshift=0cm]
    % Color-coded Header: AB in Orange
    \node[labelnode, font=\sffamily\large\bfseries] at (1,4.0) {\textcolor{orange!90!black}{AB}-type};
    \node[labelnode] at (1,3.6) {\scriptsize (ABA / ABB Avg.)};
    
    % Continuum (Positioned to reflect larger Eb ~ 0.216 eV)
    \fill[black!10] (0,2.8) rectangle (2,3.2);
    \draw[dashed, black!90] (0,2.8) -- (2,2.8);
    \node[font=\sffamily\tiny, black!100] at (1,3.0) {QP CONTINUUM};
    
    % ILX Level
    \draw[level, blue!80!black] (0,1.9) -- (2,1.9);
    \node[blue!80!black, font=\sffamily\scriptsize\bfseries] at (1,2.15) {ILX};
    
    % Eb Dimension Bracket
    \draw[dim] (-0.2, 2.8) -- (-0.2, 1.9) node[midway, left, black!100, font=\sffamily\small]{\textcolor{blue}{$E_b \approx 0.22$ eV}};
    
    % Energy Arrow
    \draw[arrow] (1,-0.2) -- (1,1.9);
    \node[labelnode, fill=white, inner sep=1.5pt, draw=green!90!black, thin, rounded corners=2pt] at (1,0.8) {\footnotesize 1.339 eV};
    
    % Dipole
    \draw[->, thick, orange!100] (-0.9, 0.2) -- (-0.9, 2.0) node[midway, left, font=\sffamily\small]{$P_z \uparrow$};
\end{scope}

% ==============================================================================
% --- THE 36 MEV SHIFT (Perpendicular Indicator) ---
% ==============================================================================
% Vertical dashed markers linking ILX levels
\draw[dashed, black!90] (2, 1.9) -- (4.5, 1.9);
\draw[dashed, black!90] (4.5, 2.26) -- (6.5, 2.26);

% Perpendicular-style shift arrow
\draw[<->, thick, red!90!black] (3.25, 1.9) -- (3.25, 2.26) 
    node[midway, right, xshift=2pt, font=\sffamily\bfseries\small]{$\Delta E \approx 36$ meV};

% ==============================================================================
% --- BA-TYPE (RIGHT) ---
% ==============================================================================
\begin{scope}[xshift=4.5cm]
    % Color-coded Header: BA in Blue
    \node[labelnode, font=\sffamily\large\bfseries] at (1,4.0) {\textcolor{blue!80!black}{BA}-type};
    \node[labelnode] at (1,3.6) {\scriptsize (BAA / BAB Avg.)};
    
    % Continuum (Positioned to reflect smaller Eb ~ 0.170 eV)
    \fill[black!10] (0,2.66) rectangle (2,3.06);
    \draw[dashed, black!90] (0,2.66) -- (2,2.66);
    \node[font=\sffamily\tiny, black!100] at (1,2.86) {QP CONTINUUM};
    
    % ILX Level
    \draw[level, blue!80!black] (0,2.26) -- (2,2.26);
    \node[blue!80!black, font=\sffamily\scriptsize\bfseries] at (1,2.51) {ILX};
    
    % Eb Dimension Bracket
    \draw[dim] (2.2, 2.66) -- (2.2, 2.26) node[midway, right, black!100]{\scriptsize \textcolor{blue}{$E_b \approx 0.17$ eV}};

    % Energy Arrow
    \draw[arrow] (1, 0.16) -- (1, 2.26);
    \node[labelnode, fill=white, inner sep=1.5pt, draw=green!60!black, thin, rounded corners=2pt] at (1, 1.21) {\footnotesize 1.375 eV};
    
    % Dipole
    \draw[->, thick, blue!100] (2.7, 2.3) -- (2.7, 0.5) node[midway, right, font=\sffamily\small]{$P_z \downarrow$};
\end{scope}

\end{tikzpicture}
\caption{Schematic representation of the energy level shifts for ILXs in AB- and BA-type stackings. Energy levels are based on EVGW$_0$ + $\mathrm{BSE}$ calculations. The diagram explicitly displays the stacking-dependent modulation of the exciton binding energy ($E_b$), showing a larger $E_b$ for AB-type ($\sim 0.22\text{ eV}$) than for BA-type ($\sim 0.17\text{ eV}$) configurations. The $\Delta E \approx 36~\mathrm{meV}$ shift represents the calculated registry-dependent offset; experimental PL energies, which represent ensemble averages, are found to be consistent with this framework \cite{schwandt-krause26_214}.}
\label{fig:BSE_Final_v3}
\end{figure}

The microscopic origin of this shift is schematically illustrated in Fig.~\ref{fig:BSE_Final_v3}. In the AB-type stacking, the internal polarization establishes a reference electrostatic potential landscape for ILX formation. Upon switching to the BA-type configuration, the reversal of the out-of-plane dipole ($P_z$) modifies the electrostatic potential profile across the layers. A closer examination reveals that this shift arises from a cooperative interplay between modest quasiparticle gap variations and the systematic $E_b$ modulations established across registries, rather than a purely rigid band-edge shift.

The predicted shift of 36~meV is in excellent agreement with the experimentally reported value of approximately 40~meV \cite{schwandt-krause26_214}, confirming that stacking-governed internal dipoles and their subsequent impact on the excitonic landscape are the dominant mechanisms for excitonic energy tuning. The remaining discrepancy ($\sim 10\%$) is likely attributable to additional dielectric screening effects, such as hBN encapsulation in experimental samples, which are not explicitly included in the present calculations.

Finally, the full range of spectral tunability reaches 108~meV between the ABA and BAB configurations, demonstrating that the trilayer architecture significantly expands the accessible energy landscape compared to conventional bilayers. By switching the stacking registry between AB- and BA-type configurations, discrete and reversible shifts in ILX emission energy can be achieved, effectively stabilizing excitons at distinct energy levels.

To bridge the gap between our first-principles findings and practical device architectures, the demonstrated multi-state energetic hierarchy and stacking-dependent polarization provide a direct physical foundation for sliding ferroelectric and optoelectronic memory concepts. Specifically, the existence of multiple stable local minima with distinct relative energies ($\Delta E$ ranging from nearly degenerate ground states to accessible metastable states up to $\sim 75\text{ meV}$, Table~S4) enables multi-state non-volatile memory, where external electrical fields or mechanical shear can deterministically switch between discrete polarization states ($P_z \uparrow$, $P_z \downarrow$). Furthermore, the robust type-II band alignment coupled with the stacking-dependent interlayer exciton energy shift ($\Delta E_{\text{ILX}} \approx 36\text{ meV}$) and distinct optical oscillator strengths allows for non-destructive optical read-out of the ferroelectric polarization state. This synergistic interplay between switchable dipole fields and tunable excitonic emission underpins the practical realization of multi-state optoelectronic memory cells and reconfigurable van der Waals heterostructure devices.

%--------------------------------------------------------------------------------------------------
%-----------------------------------------------------------------------------------------------
\section{Conclusion}
%----------------------------------------------------------------------------------------------
In summary, this work provides a comprehensive  investigation of the stacking-dependent electronic and excitonic landscapes in MoS$_2$-based vdW heterostructures. By employing eigenvalue-self-consistent $GW_0$ calculations combined with the Bethe-Salpeter equation ($GW$+BSE), we elucidate the fundamental mechanisms by which atomic registry governs interfacial polarization and quasiparticle band alignments.

Our analysis of the MoS$_2$ homobilayer demonstrates that lateral sliding between AB and BA registries enables a reversible inversion of the out-of-plane dipole moment. This switchability, driven by registry-induced symmetry breaking and asymmetric orbital hybridization, establishes the microscopic foundation for sliding ferroelectricity. In contrast, the MoS$_2$/MoSe$_2$ heterobilayer exhibits a robust, non-switchable polarization pinned by the intrinsic chemical potential mismatch between sulfur and selenium, which effectively overrides the geometric registry and stabilizes a type-II band alignment.

In trilayer 2L-MoS$_2$/MoSe$_2$ configurations, an expanded parameter space for deterministic control of charge carriers emerges. The stacking sequence dictates whether photogenerated electrons are localized in the central or bottom MoS$_2$ layer, while holes remain confined to the MoSe$_2$ monolayer. This layer-selective behavior can be described within an electrostatic framework, where internal electric fields induce band-edge shifts of 60--70~meV, directly linking dipole orientation to electronic localization.

The many-body excitonic spectra further corroborate these findings. Interlayer excitons exhibit strong sensitivity to atomic registry, with stacking-dependent energy shifts reaching up to 108~meV. Notably, the calculated average interlayer excitonic shift of 36~meV shows excellent agreement with the experimentally reported $\sim$40~meV ferroelectric shift in 3R-MoS$_2$/MoSe$_2$ heterostructures. This quantitative agreement confirms that stacking-governed internal dipoles are the primary drivers of excitonic tunability.

Overall, these results demonstrate that the interplay between structural sliding and many-body interactions in TMD trilayers provides a powerful degree of freedom for engineering programmable light--matter interactions. The ability to modulate excitonic energies and carrier transfer pathways through precise control of stacking sequence offers a promising route toward multi-state optoelectronic memory and next-generation sliding ferroelectric devices.

% ---------------------
% ACKNOWLEDGMENTS
% ---------------------
\section*{Acknowledgments}
This work was funded by the Deutsche Forschungsgemeinschaft (DFG, German Research Foundation) - SFB 1477 "Light-Matter Interactions at Interfaces", project number 441234705”. 

% ---------------------
% REFERENCES
% ---------------------
% Delete \section*{References} as natbib adds it automatically
\bibliographystyle{iopart-num}
\bibliography{ref}
\newpage % Ensure a page break before the attachment
\includepdf[pages=-]{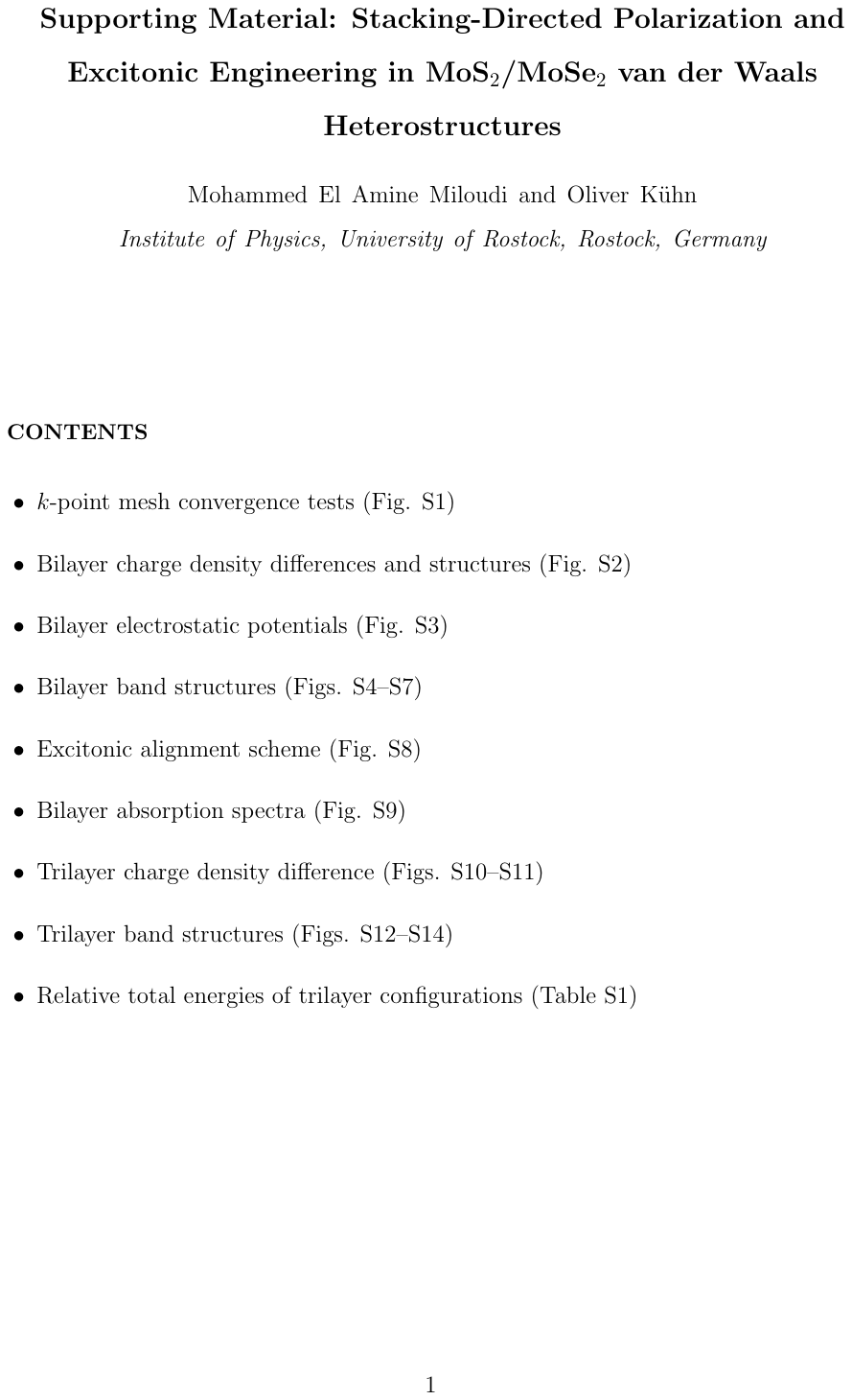}

\end{document}